\documentclass[12pt]{article}
\pdfoutput=1
\usepackage[a4paper]{geometry}
\usepackage{jheppub, amsmath,amssymb,amsfonts,amsxtra, mathrsfs, makeidx,graphics,graphicx,amsthm,epsfig, ytableau,bm,longtable,float, color,tikz,mathtools,xfrac,footnote,rotating, lscape, makecell, environ,mathtools, empheq}

\usetikzlibrary{positioning}
\usetikzlibrary{chains}
\usetikzlibrary{arrows,fit,decorations.pathreplacing}
\tikzstyle{every picture}+=[remember picture]
\tikzstyle{na} = [baseline]

\usetikzlibrary{arrows, decorations.markings, calc, fadings, decorations.pathreplacing, patterns, decorations.pathmorphing, positioning}

\def\node#1#2{\overset{#1}{\underset{#2}{{\color{gray} \bullet}}}}

\def\node#1#2{\overset{#1}{\underset{#2}{\circ}}}

\tikzstyle{every picture}+=[remember picture]
\tikzstyle{na} = [baseline=-.5ex]

\usetikzlibrary{arrows,shapes,positioning}
\usetikzlibrary{decorations.markings}
\tikzstyle arrowstyle=[scale=1.5]

\tikzstyle sdirected=[postaction={decorate,decoration={markings,
    mark=at position .52 with {\arrow[arrowstyle]{>}}}}]
\tikzstyle reverse sdirected=[postaction={decorate,decoration={markings,
    mark=at position .52 with {\arrowreversed[arrowstyle]{<};}}}]
\tikzstyle ddirected=[postaction={decorate,decoration={markings,
    mark=at position .52 with {\arrow[arrowstyle]{>>}}}}]
\tikzstyle reverse ddirected=[postaction={decorate,decoration={markings,
    mark=at position .52 with {\arrowreversed[arrowstyle]{<<};}}}]

\tikzstyle bif=[postaction={decorate,decoration={markings,
    mark=at position .4 with {\arrow[arrowstyle]{<};},mark=at position .6 with {\arrow[arrowstyle]{>};}}}]

\def\a{\alpha}
\def\b{\beta}
\def\g{\gamma}

\def\o{\omega}
\def\D{\Delta}
\def\Tr{\text{Tr}}
\def\BPS{\text{BPS}}
\def\be{\begin{equation}}
\def\ee{\end{equation}}

\title{Entropy function from toric geometry}

\author[a]{Antonio Amariti,} 
\author[b,c]{Ivan Garozzo,} 
\author[b,c]{Gabriele Lo Monaco,} 
\affiliation[a]{INFN, Sezione di Milano, Via Celoria 16, I-20133 Milano, Italy}
\affiliation[b]{Dipartimento di Fisica, Universit\`a di Milano-Bicocca, \\ Piazza della Scienza 3, I-20126 Milano, Italy}
\affiliation[c]{INFN, sezione di Milano-Bicocca, \\Piazza della Scienza 3, I-20126 Milano, Italy}
\emailAdd{antonio.amariti@mi.infn.it}
\emailAdd{ivangarozzo@gmail.com}
\emailAdd{g.lomonaco1@campus.unimib.it}
\abstract{
It has recently been claimed that a Cardy-like limit of the superconformal index
of 4d $\mathcal{N}=4$ SYM accounts for the entropy function, whose Legendre transform
corresponds to the entropy of the holographic  dual AdS$_5$ rotating black hole.
Here we study this Cardy-like limit for $\mathcal{N}=1$ toric quiver gauge theories, 
observing that the corresponding entropy function 
can be interpreted in terms of the toric data.
Furthermore, for some families of models, we compute the Legendre transform of the 
entropy function, comparing with similar results recently discussed in the literature.
}

\begin{document}
\maketitle

\section{Introduction}

The possibility of counting black hole microstates using the CFT dual picture 
is one of the most attractive consequences of the AdS/CFT correspondence 
\cite{Strominger:1996sh}.
A recent result in this field is the relation between 
the entropy of AdS$_5$ rotating black holes \cite{Gutowski:2004yv,Gutowski:2004ez,Chong:2005da,Chong:2005hr,Kunduri:2006ek}
and the superconformal index (SCI) \cite{Kinney:2005ej,Romelsberger:2005eg}.
The black hole entropy is given by the Benekstein-Hawking formula,
$S_{BH} = \frac{A}{4 G_5}$, being $A$ the area of the black hole horizon
and $G_5$ the five dimensional Newton constant.
The problem has been for a long time how to take into account 
the gravitational exponential growing ensemble of states from the dual CFT
perspective.
The potential candidate, the SCI, corresponding to the partition function 
computed on the conformal boundary $S^3 \times \mathbb{R}$,
led to a puzzle: the presence of the operator $(-1)^F$ 
induces a huge amount of cancellations between the 
bosonic and the fermionic states contributing to the index, 
and the final result is order $\mathcal{O}(1)$ instead of 
the expected $\mathcal{O}(N^2)$ \cite{Kinney:2005ej}. 

A breakthrough in the analysis has been recently given by \cite{Hosseini:2017mds}, 
where the authors associated the black hole entropy to a CFT extremization problem.
They focused on  the maximally supersymmetric case
with two angular momenta and three conserved global charges.
By reformulating the problem in term of a grand canonical
BPS partition function, $Z_{\BPS}$, they obtained the black hole entropy as a Legendre 
transform of the logarithm of $Z_{\BPS}$ (as in the cases of \cite{Strominger:1996sh,Sen:2008vm,Sen:2009vz}).
Furthermore in \cite{Cabo-Bizet:2018ehj} it was realized that $Z_{\BPS}$ can be obtained on the gravitational 
side by considering the complexified on-shell action.
% regularized at zero temperature. 

The problem of this approach is that a concrete proposal
for such a BPS partition function  is still lacking on the  field theory side.
However it is possible to perform some explicit 
calculations based on a different partition function, that can be obtained by manipulating 
the SCI \cite{Choi:2018hmj}. 
The problem of the huge cancellations between the bosonic and the fermionic states
has been circumvented by considering complex fugacities.
As discussed in \cite{Choi:2018hmj} indeed the cancellations  
are \emph{optimally  obstructed} by the imaginary parts of the fugacities at the saddle point.
The existence of a deconfinement transition in presence of 
complex fugacities was then observed in \cite{Choi:2018vbz}.
Moreover the authors of \cite{Benini:2018ywd} 
exploited a reformulation of the SCI of 
$4d$ $\mathcal N=1$ theories as a finite sum over the solution of the 
so-called Bethe Ansatz equation \cite{Benini:2018mlo}.

Using these ideas the authors of 
\cite{Benini:2018ywd,Choi:2018hmj,Honda:2019cio,ArabiArdehali:2019tdm,Kim:2019yrz,Cabo-Bizet:2019osg} have obtained the 
BPS entropy function $S_E(\Delta_1,\Delta_2,\Delta_3,\omega_1,\omega_2)$ of \cite{Hosseini:2017mds} from the SCI of $\mathcal{N}=4$ SYM. At large $N$ this function reads 
\begin{equation}
S_E = -i \pi N^2 \frac{\Delta_1 \Delta_2 \Delta_3 }{\omega_1 \omega_2}
\end{equation}
where $\Delta_I$ and $\omega_a$ are the fugacities
conjugated to the charges $Q_I$ and $J_a$ of the 
$SO(6)_R$ R-symmetry and the $SO(4) \subset SO(4,2)$ 
conformal symmetry respectively.
Furthermore these fugacities are constrained by the relation
\footnote{See \cite{Cabo-Bizet:2018ehj} for a detailed explanation on the sign.} 
 $\Delta_1+\Delta_2+\Delta_3-  \omega_1-\omega_2 = 1$,  
that corresponds, on the supergravity dual, to a stability condition 
on the killing spinor \cite{Cabo-Bizet:2018ehj}.

A natural question regards the extension of this result to other families 
of 4d $\mathcal{N}=1$ SCFT with an holographic dual description.
Recent attempts in this direction has been given in \cite{Honda:2019cio}, for 
the case of necklace $\mathcal{N}=2$ models, in \cite{Kim:2019yrz} for the case of 
$Y^{pp}$ family and in \cite{Cabo-Bizet:2019osg} for more general classes of superconformal quivers.
In all these cases the authors considered a subgroup of the full global symmetry 
and found interesting extensions of the results, showing also that the 
Legendre transform led to the expected entropy of the dual black hole.

In this paper we focus on infinite families of models, denoted as toric quiver gauge theories,
that include the cases considered so far. These models  describe the low energy dynamics of a stack 
of $N$ D3 branes probing the tip of a toric cone over a five dimensional
Sasaki-Einstein manifold.
We study the large $N$ index in the Cardy-like limit with complex fugacities discussed above
and we give evidences of a general relation of the form 
\begin{equation}
\label{gen}
S_E = -i \pi N^2 \frac{ C_{IJK} \Delta_I \Delta_J \Delta_K }{6\,\omega_1 \omega_2}
\end{equation}
where the fugacities $\Delta_I$ are read from the toric data and they satisfy 
the constraint $\sum_{I=1}^{d} \Delta_I - \sum_{a=1}^{2} \omega_a  = 1$.
This results has been already conjectured in \cite{Hosseini:2018dob,Zaffaroni:2019dhb},  
where it was proposed that the  numerator of (\ref{gen}) has the functional structure of 
the conformal anomaly  of the 4d theory extracted from the gravitational (or geometric) data.
The coefficients $C_{IJK}$ in  (\ref{gen})  corresponds to the Chern-Simons couplings
of the holographic dual gravitational description. 
Under the AdS/CFT correspondence they are associated to the triangle anomalies 
of the SCFT  as shown in \cite{Benvenuti:2006xg}.

The paper is organized as follows.
In section \ref{Sec:Cardy} we review the main aspects of our calculation 
focusing on the Cardy-like limit of the superconformal index
and on the relation between the toric data and the global symmetries of the dual
field theory.
In section \ref{Sec:conifold} we study the case of the conifold, computing the Cardy-like limit of 
the SCI and giving some evidences for the general conjecture on the 
behaviour of the gauge holonomies at the saddle point.
In section \ref{Sec:OtherExamples} we study other simple examples of toric quiver gauge theories,
showing the validity  of (\ref{gen}) for each case.
In section \ref{Sec:Families} we focus on some infinite families, $Y^{pq}$, $L^{pqr}$ and $X^{pq}$ theories, and also 
in these cases we give evidences of (\ref{gen}).
In section \ref{Sec:Legendre} we discuss the Legendre transform of the formula for the entropy of $\mathcal{N}=2$ necklace 
quivers and for quivers in the $Y^{pp}$ family.
In both cases we extend the results already computed in the literature by turning on all the global symmetries.
In section \ref{Sec:Conclusions} we conclude, discussing possible future lines of research.

\section{The Cardy-like  limit of toric quivers}
\label{Sec:Cardy}

In this section we explain the general aspects of the calculation of the
Cardy-like limit of the SCI with complex fugacities
for toric quiver gauge theories.

Toric quiver gauge theories  describe the low energy dynamics of a stack 
of $N$ D3 branes probing the tip of a toric cone over a five dimensional
Sasaki-Einstein manifold.
The toric data describing the singularity can be associated with  
the field theory data obtained by studying the moduli space
\cite{Hanany:2005ve,Franco:2005rj}.
In order to obtain these data starting from a gauge theory one has
to first embed the quiver in a two dimensional torus.
In this way one obtains a planar diagram, that can be transformed in 
a dimer, by exchanging faces and nodes.
On this structure one defines the notion of perfect matching (PM): the PMs are
collections of fields that represent all the possible dimer covers.  
By weighting the PMs with respect to the one-cycles of the first homology group of
the torus one defines two possible intersection numbers for each PM.
One can then assign a vector $V_I = (\cdot,\cdot,1)$ to each PM, such that 
the first two entries are the intersection numbers discussed above and the last one is fixed to 1.
The toric diagram corresponds to the convex integral polygon constructed from the $V_I$ vectors.
Using this construction it is possible to assign a basis of global symmetries
of the quiver directly from the toric diagram.
This consists of assigning a  $U(1)_I$ symmetry, denoted as $Q_I$, to each external point of the toric diagram.
One can construct the $R$-symmetry and the flavor (and baryonic symmetries) by combining these $U(1)_I$ as follows.

First one assigns a set of coefficients $\bf{a}_I \equiv \{{\bf a}_I^{(R)},{\bf a}_I^{(i)}\}$ to each PM.
Then it is necessary to impose the constraints $\sum_{I=1}^{d} {\bf a}_I^{(R)}=2$ and  $\sum_{i=1}^{d} {\bf a}_I^{(i)}=0$
$\forall i $, where $d$ is the number of external points in the toric diagram.
The charges of the fields are associated to the ones of the PM with the prescription of \cite{Butti:2005vn}.
Furthermore, the areas  of the  triangles obtained by connecting three external points  of the
toric diagram coincide with the triangular anomalies between the three 
$U(1)_I$ symmetries associated to such points
\cite{Benvenuti:2006xg}
\begin{equation}
\label{BPT}
\frac{N^2}{2} |\det (V_I, V_J, V_K)|  = \Tr(Q_I Q_J Q_K) \equiv N^2 C_{IJK}
\end{equation}

As an example let us discuss the simplest toric quiver gauge theory, corresponding to $\mathcal{N}=4$ 
$SU(N)$ SYM.
We look at this theory as an $\mathcal{N}=1$ theory with superpotential
\begin{equation}
W = \Phi_1 [\Phi_2,\Phi_3]
\end{equation} 
where $\Phi_I$ are in the adjoint gauge group.
In this case we have three $U(1)$ trial R-symmetries,
denoted as $2 U(1)_{1,2,3}$, and each $U(1)_I$ assigns charge $1$ 
to the $I$-th field and zero to the others:
\begin{equation}
\begin{array}{c|ccc}
& U(1)_1& U(1)_2& U(1)_3 \\
\hline
\Phi_1&1&0&0 \\
\Phi_2&0&1&0 \\
\Phi_3&0&0&1 \\
\end{array}
\end{equation}
There are three PM as shown in (\ref{PMS5}), corresponding to the three fields $\Phi_I$.
\be 
\label{PMS5}
\scalebox{1}
{
\begin{tikzpicture}[font=\scriptsize, baseline] \begin{scope}[auto,%
  every node/.style={draw}, node distance=1cm]; 

\node[draw=none] at (-0.95, 1.4) {\tiny{1}};
\node[draw=none] at (0, 2) {\tiny{1}};
\node[draw=none] at (0.95, 1.4) {\tiny{1}};
\node[draw=none] at (0, 0.8) {\tiny{1}};
\end{scope}

\draw[-o](-0.6,1.4)--(0.7,1.4);
 \draw  node[fill,circle,scale=0.4]at (-0.6,1.4) {};
\draw (-0.97,1.83)--(-0.62,1.48);
\draw (0.97,0.97)--(0.62,1.33);
\draw[line width=0.4mm, blue]  (0.97,1.83)--(0.62,1.48);
\draw[line width=0.4mm, blue] (-0.97,0.97)--(-0.62,1.32);

\draw[thick,red] (0,0)--(-1.4,1.4);
\draw[thick,red] (0,0)--(1.4,1.4);
\draw[thick,red] (-1.4,1.4)--(0,2.8);
\draw[thick,red] (0,2.8)--(1.4,1.4);

\end{tikzpicture}} \qquad 
\scalebox{1}
{
\begin{tikzpicture}[font=\scriptsize, baseline] \begin{scope}[auto,%
  every node/.style={draw}, node distance=1cm]; 
\node[draw=none] at (-0.95, 1.4) {\tiny{1}};
\node[draw=none] at (0, 2) {\tiny{1}};
\node[draw=none] at (0.95, 1.4) {\tiny{1}};
\node[draw=none] at (0, 0.8) {\tiny{1}};

\end{scope}

\draw[-o](-0.6,1.4)--(0.7,1.4);
 \draw  node[fill,circle,scale=0.4]at (-0.6,1.4) {};

\draw(0.97,1.83)--(0.62,1.48);
\draw (-0.97,0.97)--(-0.62,1.32);

\draw[line width=0.4mm, blue]   (-0.97,1.83)--(-0.62,1.48);
\draw[line width=0.4mm, blue]   (0.97,0.97)--(0.62,1.325);
\draw[thick,red] (0,0)--(-1.4,1.4);
\draw[thick,red] (0,0)--(1.4,1.4);
\draw[thick,red] (-1.4,1.4)--(0,2.8);
\draw[thick,red] (0,2.8)--(1.4,1.4);

\end{tikzpicture}} \qquad 
\scalebox{1}
{
\begin{tikzpicture}[font=\scriptsize, baseline] \begin{scope}[auto,%
  every node/.style={draw}, node distance=1cm]; 
\node[draw=none] at (-0.95, 1.4) {\tiny{1}};
\node[draw=none] at (0, 2) {\tiny{1}};
\node[draw=none] at (0.95, 1.4) {\tiny{1}};
\node[draw=none] at (0, 0.8) {\tiny{1}};

\end{scope}

\draw[-o](-0.6,1.4)--(0.7,1.4);
 \draw  node[fill,circle,scale=0.4]at (-0.6,1.4) {};
\draw[line width=0.4mm, blue]  (-0.52,1.4)--(0.55,1.4);
\draw (-0.97,1.83)--(-0.62,1.48);
\draw (0.97,0.97)--(0.62,1.33);
\draw (0.97,1.83)--(0.62,1.48);
\draw (-0.97,0.97)--(-0.62,1.32);

\draw[thick,red] (0,0)--(-1.4,1.4);
\draw[thick,red] (0,0)--(1.4,1.4);
\draw[thick,red] (-1.4,1.4)--(0,2.8);
\draw[thick,red] (0,2.8)--(1.4,1.4);

\end{tikzpicture}}
\ee
The toric diagram is then generated by the three vectors
\begin{equation}
\label{vectors}
V_1 = (0,0,1),
\quad
V_2 = (0,1,1),
\quad
V_3 = (1,0,1)
\end{equation}
The three trial $R$-symmetries are associated to the three corners
of the toric diagram generated by three vectors in  (\ref{vectors}).
Combining these symmetries we can extract the $U(1)_R$ symmetry 
and the other two flavor symmetries associated to the Cartan of the $SU(4)_R$
symmetry group of $\mathcal{N}=4$ SYM. 
For example we can choose as an $R$-symmetry the combination
$\frac{2}{3} (U(1)_1+U(1)_2+U(1)_3)$. 
In this case this assigns $R$-charge $\frac{2}{3}$ to each fields and it 
gives accidentally also the exact $R$-symmetry of the model.
More generally the exact $R$-symmetry is given by $a$-maximization \cite{Intriligator:2003jj}, where the conformal anomaly 
in this language corresponds to the  function \cite{Martelli:2005tp,Butti:2005vn}
\begin{equation}
a_{\text{geom}} \propto C_{IJK} {\bf a}_{I}^{(R)}{\bf a}_{J}^{(R)}{\bf a}_{K}^{(R)}
\end{equation}
The other two global symmetries can be obtained by the combinations
$U(1)_1'=U(1)_1-U(1)_3$ and $U(1)_2' = U(1)_2-U(1)_3$.
In this way we assign the charges as
\begin{equation}
\label{basis}
\begin{array}{c|ccc}
& U(1)_1'& U(1)_2'& U(1)_R \\
\hline
\Phi_1&1&0&\frac{2}{3} \\
\Phi_2&0&1&\frac{2}{3}  \\
\Phi_3&-1&-1&\frac{2}{3}  
\end{array}
\end{equation}

Using these ideas one can read the parameterization of the 
global symmetries entering in the superconformal index 
from the toric diagram. 
We just have to linearly combine the $U(1)_I$  symmetries in order to
obtain the non-$R$, either flavor or baryonic symmetries.
In the following we will choose the $d-1$ combinations
$U(1)_i-U(1)_d$, with $i=1,\dots d-1$ as our basis of non $R$-global symmetries .
Furthermore the $R$-symmetry (not necessarily the exact one) 
will correspond to the combination $\frac{2}{d}\sum_{I=1}^{d} U(1)_I$.
Using this basis of charges and symmetries we can write the SCI
of a toric quiver gauge theory in the form
\begin{equation}
I = \text{Tr}_{\text{BPS}}  \, (-1)^F  e^{-\beta H} p^{J_1+  \frac{1}{d}  \sum_{i=1}^{d}   Q_i } q^{J_2+  \frac{1}{d}  \sum_{i=1}^{d}    Q_i } \prod_{i=1}^{d-1} {u_i}^{    Q_i -    Q_d}
\end{equation}
Then we shift the chemical potentials $u_i \rightarrow u_i (pq)^{-\frac{1}{d}}$ 
obtaining 
\begin{equation}
I = \text{Tr}_{\text{BPS}}  \, (-1)^F  p^{J_1} q^{J_2} (pq)^{\frac{1}{d}  \sum_{i=1}^{d}   Q_i  } \prod_{i=1}^{d-1} u_i^{  Q_i -    Q_d} 
(pq)^{\frac{  Q_d -   Q_i}{d}}
\end{equation}
Then by defining $p = e^{2 \pi i \omega_1}$, $q = e^{2 \pi i \omega_2}$, 
$u_i = e^{2 \pi i \Delta_i}$ (for $i = 1\dots d-1$)
and  $(-1)^F  = e^{2 i \pi    Q_d}$ (using the fact that this is an R-symmetry as well) we can express the index as
\begin{equation}
\label{finalSCI}
I = \Tr_{\BPS} \,  e^{2 \pi  i \omega_1 J_1 } e^{2 \pi i \omega_2 J_2 } \prod_{I=1}^{d} e^{2 \pi i \Delta_I   Q_I}
\end{equation}
with the constraint
\begin{equation}
\label{Constraint}
\sum_{I=1}^{d} {\Delta_I} - \omega_1 - \omega_2 = 1 
\end{equation}
The Cardy  limit of the SCI \cite{DiPietro:2014bca} and its generalization in 
\cite{DiPietro:2016ond,Ardehali:2015bla}
are obtained by shrinking the circle on which the index is defined as a partition 
function $S^3 \times S^1$.
This can be done with complex fugacities by taking the limit $|\omega_1|,|\omega_2| \rightarrow 0$ 
\cite{Choi:2018hmj,Honda:2019cio,ArabiArdehali:2019tdm}
\begin{equation}
\lim_{|\omega_1|,|\omega_2| \rightarrow 0} I \simeq e^{- \frac{i \pi (\omega_1+\omega_2)}{12 \omega_1\omega_2}\text{Tr} R} 
\int \prod_{i=1}^{\text{rank} \, \mathcal{G}} d a_i e^{V( a)}  
\end{equation}
where
\begin{equation}
V(a) = \frac{i \pi}{2 \omega_1\omega_2} \left( V_1(a) (\omega_1+\omega_2)+\frac{V_2(a)}{3} \right)
\end{equation}
In this formula $\text{rank}\,\mathcal{G}$ refers to the dimension of the maximal abelian torus of 
the gauge group,
that is parameterized by the gauge holonomies $e^{2 \pi i a_i}$.
The functions $V_1$ and $V_2$ are
\begin{eqnarray}
V_{1}(a)
&=&
\sum_{k=1}^{G} \sum_{m,n=1}^{N} \theta\Big(a_m^{(k)}-a_n^{(k)}\Big)
+
\sum_{k \rightarrow k'}
\sum_{m,n=1}^{N} 
(R_{kk'}-1)\theta\Big(
a_m^{(k)}-a_n^{(k')}
+
\sum_{i=1}^{d-1}
q_{kk'}^i \Delta_i
\Big)
\nonumber \\
V_{2}(a)
&=&-
\sum_{k \rightarrow k'}
\sum_{m,n=1}^{N} 
\kappa\Big(
a_m^{(k)}-a_n^{(k')}
+
\sum_{i=1}^{d-1}
q_{kk'}^i \Delta_i
\Big)
\end{eqnarray}
Let us explain these formulas. In the first line 
$G$ refers to the number of gauge groups.
It is obtained from a toric diagram by 
the formula $G= 2\mathcal{I} + d - 2$, where $\mathcal{I} $ 
is the number of internal points.
In the formula for $V_1$ there are two contributions, the
first comes from the vector multiplets while the second from each bifundamental
multiplet connecting the $k$-th to the $k'$-th node.
Adjoints matter fields have $k = k'$.
The function $V_2$ takes contributions only from the matter fields.
Each matter field has $R$-charge $R_{kk'}$ and global charges
$q_{kk'}^i$.  The fugacities $\Delta_i$ are the ones defined above.
In this paper we will always refer to $SU(N)$ gauge theories, and this will impose the constraint 
 $\sum_{i=1}^N a_m ^{(k)}=0$.
  Moreover, the functions 
$\theta(x)$ and $\kappa(x)$ are given by
\begin{equation}
	\theta(x)=\{x\}(1-\{x\}), \qquad \kappa(x)=\{x\}(1-\{x\})(1-2\{x\}),
\end{equation}
with the fractional part $\{x\}=x-[x]$, and can be rewritten as
\begin{equation}
\label{simpk}
	\theta(x)=|x|-x^2, \qquad \kappa(x)=2x^3-3x|x|+x
\end{equation}
for $|x| \le 1$.
The next step consists of evaluating the integral (\ref{finalSCI}). 
We start by ignoring the contribution of $\Tr \,R$. This is because in this paper
we  always consider  toric quivers
with a weakly coupled gravity dual. It follows that the gravitational anomaly,
proportional to $\Tr \, R$,  is order $\mathcal{O}(1)$, while we restrict to 
the leading large $N$ contribution of the
Cardy-like limit of the index.
Furthermore, we focus on the regime 
Re$\left(\frac{i}{\omega_1 \omega_2 }\right) >0$. 
This is the regime discussed in \cite{Choi:2018hmj,Honda:2019cio,ArabiArdehali:2019tdm,Kim:2019yrz,Cabo-Bizet:2019osg}
where it was shown that there is a saddle point at vanishing holonomies 
when considering $\mathcal{N}=4$ SYM. 
Computing the Cardy-like limit of  the SCI 
using the charges (\ref{basis}) we obtain the 
entropy function 
\begin{equation}
\label{n4res}
S_E = -i \pi N^2 \frac{\Delta_1 \Delta_2 \Delta_3 }{\, \omega_1 \omega_2}
\end{equation}
where $\Delta_I$  are the 
fugacities of the symmetries $U(1)_I$ with $I=1,2,3$ and  
 $\Delta_1+\Delta_2+\Delta_3 - \omega_1 - \omega_2 = 1$.

Here we study more general classes of quiver gauge theories.
The first problem corresponds to find arguments in favor of
the existence on an universal saddle point with vanishing holonomies
as already discussed in 
\cite{Choi:2018hmj,Honda:2019cio,ArabiArdehali:2019tdm,Cabo-Bizet:2019osg}.
Here we will confirm this expectations, observing in  examples on increasing complexity that 
there is always a regime of fugacities  that allows the
existence of such a universal saddle.

Furthermore in each example we compute the Cardy-like limit of the index 
at large $N$, and we observe that 
it is controlled by the function
\begin{equation}
\label{expected}
S_E = -i \pi N^2 \frac{C_{IJK} \Delta_I \Delta_J \Delta_K }{6 \, \omega_1 \omega_2}
\end{equation}
where $\Delta_I$ are  the fugacities appearing in (\ref{finalSCI})  and the constraint 
(\ref{Constraint}) is imposed.
This result can be proved by considering  the relation obtained in 
\cite{Cabo-Bizet:2018ehj,Kim:2019yrz} for the Cardy-like limit of the SCI of a generic 
$\mathcal{N}=1$ gauge theory in presence of flavor fugacities.
The relation is 
\begin{equation}
\label{TrRF}
S_E =-i \pi N^2  \frac{\Tr(\Delta R  + x_i F_i)^3}{6\,\omega_1 \omega_2} 
\end{equation}
that holds imposing  the constraint 
\begin{equation}
\label{constr}
2\Delta - \omega_1-\omega_2 =1
\end{equation} 
where $\Delta$ represents the $R$-symmetry fugacity,
while $x_i$ are the flavor symmetry fugacities.
We can express the $R$-symmetry and the flavor symmetries $F_i$ as 
%combinations of the generators of the $Q_I$ $U(1)$  symmetries introduced above.
%We have
\begin{equation}
R = \sum_{I=1}^{d}Q_I\,{\bf a}_I^{(R)}, 
\quad
F_i = \sum_{I=1}^{d} Q_I\,{\bf a}_I^{(i)} \quad \forall i
\end{equation}
with the constraints $\sum_{I=1}^{d} {\bf a}_I^{(R)}=2$ and $\sum_{I=1}^{d} {\bf a}_I^{(i)}=0$, $\forall i$.
The combination appearing in (\ref{TrRF}) can be expressed in terms of these redefinitions as
\begin{equation}
\Delta R + x_i F_i 
=
\sum_{I=1}^{d} Q_I \bigg( \Delta \, {\bf a}_I^{(R)} + \sum_{i=1}^{d-1} x_i {\bf a}_I^{(i)} \bigg)
\equiv
\sum_{I=1}^d  Q_I \Delta_I
\end{equation}
where in the last equality we defined the new fugacities
$\Delta_I$.
These fugacities are constrained as
\begin{equation}
\label{newconstr}
\sum_{I=1}^{d} \Delta_I 
= 
\sum_{I=1}^{d}  \bigg( \Delta \, {\bf a}_I^{(R)}+ \sum_{i=1}^{d-1} x_i {\bf a}_I^{(i)} \bigg)
=
2\Delta = \omega_1+\omega_2 +1
\end{equation}
where in the last equality we used the constraint (\ref{constr}).
In terms of the $Q_I$ symmetries the entropy function reads  
\begin{eqnarray}
\label{final}
S_E  
&=&
-\frac{i \pi N^2}{6\,\omega_1 \omega_2} \text{Tr} \Big(\sum_{I=1}^d  Q_I \Delta_I \Big)^3
=
-\frac{i \pi N^2}{6\,\omega_1 \omega_2} \text{Tr}( Q_I Q_J Q_K) \Delta_I \Delta_J \Delta_K
\nonumber \\
&=&
-\frac{ i \pi N^2}{6\,\omega_1 \omega_2} C_{IJK} \Delta_I \Delta_J \Delta_K
\end{eqnarray}
with the constraint (\ref{newconstr}) and 
the last equality follows from the relation (\ref{BPT}). 
We are going to verify (\ref{final}) in the rest of the paper by explicitly studying the Cardy-like limit of the SCI for 
many toric quiver gauge theories.

\section{The conifold}
\label{Sec:conifold}
The conifold represents an ideal arena where testing, at finite rank, the Cardy formula and show the agreement with our general proposal \eqref{expected}, once the charges are parametrized from a geometric point of view. 

The theory we are going to study has been proposed originally in \cite{Klebanov:1998hh} as the theory living on a stack of $N$ D3-branes probing the tip of the conical singularity $xy-zt=0$ ; taking the near-horizon limit, the theory turns out to be holographically dual to $\text{AdS}_5\times T^{1,1}$ background where $T^{1,1}$ is what is properly named conifold. $T^{1,1}$ can be seen as an $U(1)$ fibration over $\mathbb{CP}^1\times\mathbb{CP}^1$ with the $U(1)$ fiber playing the role of Reeb vector; the manifold admits a Sasaki-Einstein structure and has the topology of $S^2\times S^3$
More importantly for our discussion, $T^{1,1}$ is also toric, with the toric diagram identified by the following four vectors:
\begin{equation}
V_1\,=\, (1,0,0),\quad V_2\,=\, (1,1,0),\quad V_3\,=\,(1,1,1)\,,\quad V_4\,=\,(1,0,1).
\end{equation}
The dual theory can be summarized by the following quiver and superpotential:
\begin{equation}
\scalebox{1}
{
\begin{tikzpicture}[font=\scriptsize, baseline] 
\tikzset{decoration={snake,amplitude=.4mm,segment length=2mm,
                       post length=0mm,pre length=0mm}}
\begin{scope}[auto,%
  every node/.style={draw, minimum size=1.2cm}, node distance=1cm]; 
 \node[draw=none] (n1) at (0,0) {$\bullet$};
  \node[draw=none] at (-0.3,0) {$1$};
 \node[draw=none] (n2) at (4,0) {$\bullet$};
  \node[draw=none] at (4.3,0) {$2$};
\end{scope}
%\draw[black] (node1) edge [out=-45,in=-135,loop,looseness=4] node[midway,below=-0.2cm] {$\phi_1$}  (node1);%
\draw[black] (0,0) edge [out=30,in=150, ddirected] node[midway,above=0.1cm] {$A^1,A^2$}  (4,0);
\draw[black] (4,0) edge [out=210,in=-30, ddirected] node[midway,below=0.1cm] {$B^1,B^2$} (0,0);
\end{tikzpicture}
}
\quad\quad
W\,\propto\, \epsilon_{ij}\epsilon_{kl}\,\text{Tr}\left(A^iB^kA^jB^l\right)
\end{equation}
The isometries of $T^{1,1}$ suggest the global symmetries of the CFT: a $U(1)_R$ factor (the $R$-symmetry generated by action of the Reeb vector) and two $SU(2)$ factors to be identified with the isometries of $\mathbb{CP}^1\times\mathbb{CP}^1$; finally, we need to add a $U(1)_B$ baryonic symmetry associated to the unique non-trivial three-cycle of the geometry\footnote{As we said, the topology of $T^{1,1}$ is actually the same of $S^2\times S^3$. The unique three-cycle can be understood as this $S^3$.}.
The charges of the fields under $U(1)_R\,,U(1)_B$ and (a combination of) the Cartan generators $U(1)_{1,2}$ of the $SU(2)$ factors are summarized in the table below:
\begin{equation}
\begin{tabular}{l | c | c| c| r}
&    $U(1)_R$    &    $U(1)_B$    &    $U(1)_1$   &    $U(1)_2$ \\
\hline
$A_1$    &   $1/2$   &   $1$   &   $1$   &  $1$ \\
$A_2$    &   $1/2$   &   $1$   &   $-1$   &  $-1$ \\
\hline
$B_1$    &   $1/2$   &   $-1$   &   $1$   &   $-1$ \\
$B_2$    &   $1/2$   &   $-1$   &   $-1$  &  $1$ \\
\end{tabular}
\end{equation}
We will turn on fugacities $\Delta_{F_{1,2}}$ for the flavour symmetries $U(1)_{1,2}$ and fugacity $\Delta_B$ 
for the baryonic symmetry $U(1)_B$. 

We want to study now the Cardy formula in the rank-1 case, {\it i.e.} for $SU(2)$ gauge groups; in fact, a crucial point is understanding the behaviour of the saddle points with respect to the holonomies. In low-rank cases it is possible to prove the main conjecture, {\it i.e.} it is possible to find charge configurations where the dominant saddle-point contribution is unique and corresponds to putting to zero all the holonomies; then, we will generalize to arbitrary $N$ assuming the conjecture to be true at any rank.
This fits with the discussions on the existence of such and universal saddle point in
 \cite{ArabiArdehali:2019tdm, Kim:2019yrz,Cabo-Bizet:2019osg}. 
Moreover, we want to show that the choice of range for the fugacities is crucial and not all of them are suitable for our purpose.
Let us start evaluating:
\begin{equation}
\label{V2conifold1}
\begin{split}
V_2\,=\,\!-\!\sum_{m,n=1}^{N} &\left(\kappa\left[a_m^{(1)}\!-\!a_n^{(2)}+\Delta_{F_1}+\Delta_{F_2}+\Delta_{B}\right]
\!+\!\kappa\left[a_m^{(1)}-a_n^{(2)}-\Delta_{F_1}\!-\!\Delta_{F_2}+\Delta_{B}\right]+\right.\\ 
&\left.\,\,\,\,\kappa\left[a_m^{(2)}\!-\!a_n^{(1)}+\Delta_{F_1}\!-\!\Delta_{F_2}-\Delta_{B}\right]\!+\!\kappa\left[a_m^{(2)}-a_n^{(1)}
-\Delta_{F_1}+\Delta_{F_2}-\Delta_{B}\right] \right)\,,
\end{split}
\end{equation}
where  $a^{(1)}_m$ and $a^{(2)}_m$ are the holonomies for the first and second gauge group respectively. In the $SU(2)$ case we also need to enforce the condition $a^{(k)}_2\,=\,-a^{(k)}_1$ so that we are actually left with just two independent variables; in the following it will be more convenient to use the  combinations:
\begin{equation}
a_{\pm}\,=\,a^{(1)}_1\,\pm\,a^{(2)}_1\,.
\end{equation}
After some algebraic manipulation, \eqref{V2conifold1} can be reduced to
\begin{equation}
V_2\,=\,-\left( f[a_+]+f[a_-] \right)\,,
\end{equation}
where the function $f$ is defined as follows:
\begin{equation}
\begin{split}
f[x]\,=\,&\kappa[x+\Delta_{F_1}+\Delta_{F_2}+\Delta_{B}]-\kappa[x-\Delta_{F_1}-\Delta_{F_2}-\Delta_{B}]\,+\\
+&\kappa[x-\Delta_{F_1}-\Delta_{F_2}+\Delta_{B}]-\kappa[x-\Delta_{F_1}-\Delta_{F_2}-\Delta_{B}]\,+\\
+&\kappa[x+\Delta_{F_1}-\Delta_{F_2}-\Delta_{B}]-\kappa[x-\Delta_{F_1}+\Delta_{F_2}+\Delta_{B}]\,+\\
+&\kappa[x-\Delta_{F_1}+\Delta_{F_2}+\Delta_{B}]-\kappa[x+\Delta_{F_1}-\Delta_{F_2}+\Delta_{B}]\,.
\end{split}
\end{equation}

Extremizing $V_2$ amounts to find extrema of $f[x]$. Observe that this function is invariant under permutations of fugacities 
$\Delta{F_1}\,,\Delta_{F_2}$ and $\Delta_{B}$. It follows that we can choose an ordering of the charges without loss of generality, let us say $0\leq \Delta_{F_1}\leq \Delta_{F_2}\leq \Delta_{B}$.
Furthermore, using the property $\kappa[x]\,=\,\kappa[x+1]$  we can move to a region where $-1/2 \leq \Delta_{F_1}+
\Delta_{F_2}+\Delta_{B} \leq 1/2$. We want to focus for simplicity on a particular ``chamber'', where we fix $0\leq \D_B, \D_{F_1},\D_{F_2}\leq1/2$; this choice almost fixes completely the chamber and an ordering for all possible combinations $\D_{F_1}\pm \D_{F_2}
\pm \D_B$. 
We are left with two possibilities:
\begin{equation}
\D_{F_1}+\D_{F_2}\geq \D_B\quad \text{or} \quad \D_{F_1}+\D_{F_2}\leq \D_B\,.
\end{equation}
Now we are able to analytically evaluate $f[x]$ in the ``fundamental regions'' $|x|<1-\D_{F_1}-\D_{F_2}+\D_B$ and $|x|<1+
\D_{F_1}+\D_{F_2}-\D_B$ respectively, where we can use the simplified expression \eqref{simpk}. We proceed to study the behaviour of $f[x]$ and we will see that these regimes are physically different and do not share the same properties.
\begin{itemize}
\item $\mathbf{\D_B\geq \D_{F_1}+\D_{F_2}}$: In this case the fundamental region is $|x|<1+\D_{F_1}+\D_{F_2}-B$ and $f[x]$ reads:
{\small
\begin{equation}
\begin{split}
&f[x] =\\ 
&\left\{
\begin{array}{lr}
 48 \Delta_{F_1} \Delta_{F_2} (2 \D_B-1)\,\,\,\, & 0<x<x_1 \\
 6 \left(\left(x+\Delta_{F_1}+\Delta_{F_2}-\D_B\right){}^2+8 (2 \D_B-1) \Delta_{F_1} \Delta_{F_2}\right) \,\,\,\, &
 x_1<x<x_2 \\
 24 \Delta_{F_1} \left(\left(4 \Delta_{F_2}-1\right) \D_B+x-\Delta_{F_2}\right) \,\,\,\,& 
x_2<x<x_3 \\
 6 \left(16 \D_B \Delta_{F_1} \Delta_{F_2}-\left(x-\D_B-\Delta_{F_1}-\Delta_{F_2}\right){}^2\right) \,\,\,\,& 
x_3<x<x_4 \\
 96 \D_B \Delta_{F_1} \Delta_{F_2} \,\,\,\,& x>x_4\\
\end{array}
\right.
\end{split}
\end{equation}}
where
$x_1=\D_B-\Delta_{F_1}-\Delta_{F_2}$,
$x_2=\D_B+\Delta_{F_1}-\Delta_{F_2}$,
$x_3 = \D_B-\Delta_{F_1}+\Delta_{F_2}$,
and
$x_4=\D_B+\Delta_{F_1}+\Delta_{F_2}$

We can observe that in a whole neighborhood of $x=0$ the function is constant; thus, for vanishing holonomies, $f[x]$ exhibits a plateaux of extrema, rather than a unique minimum or maximum, as we can see from its plot in figure \ref{ConifoldFx2}.
As a consequence, in order to evaluate the index,  we should perform an integration over the whole plateaux, making the study harder. For this reason, we exclude this case from our analysis but we will comment more about this point in the conclusions.
\begin{figure}[!h]
\centering
  \includegraphics[width=0.6\textwidth]{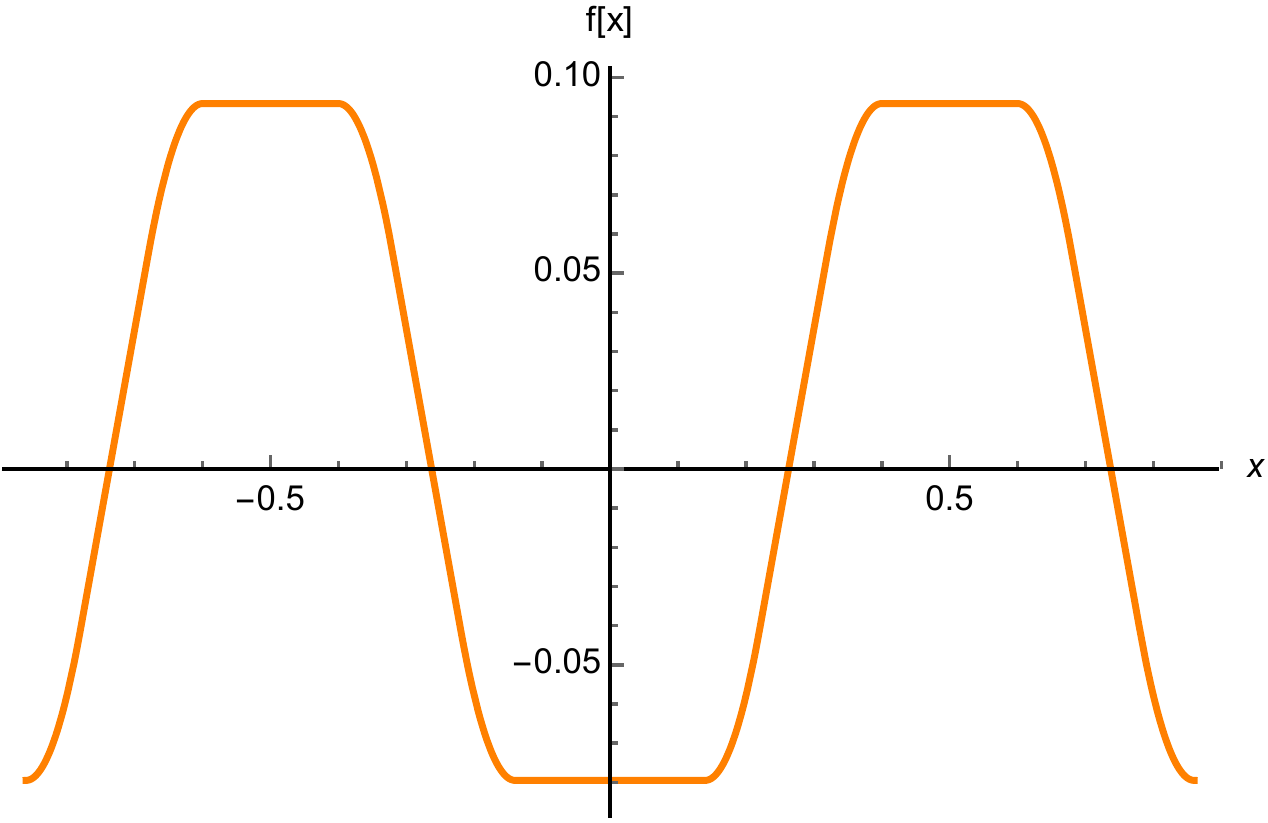}
    \caption{Plot of $f[x]$ for $\{\Delta_{F_1},\D_{F_2},\D_B\}\,=\,\{0.04\,,0.9\,,0.27\}$}
    \label{ConifoldFx2}
    \end{figure}
    
\item $\mathbf{\D_B\leq \D_{F_1}+\D_{F_2}}$: In this case the fundamental domain is $|x|<1-\D_{F_1}-\D_{F_2}+\D_B$ and $f[x]$ reads:
{\small
\begin{equation}
\begin{split}
&f[x] = \\
&\left\{
\begin{array}{lr}
 12(\D_B^2+(\D_{F_1}-\D_{F_2})^2-2\D_B\D_{F_2}+\D_B\D_{F_1}(8\D_{F_2}-2)+x^2)  
 & 
 0<x<x_1\\
 6 ((x+\D_{F_1}+\D_{F_2}-\D_B)^2-8 (1-2 \D_B) \D_{F_1} \D_{F_2}) 
 & 
 x_1<x<x_2\\
 24 \D_{F_1} (x-\D_{F_2}+\D_B (4 \D_{F_2}-1)) 
 & 
x_2<x<x_3
 \\
 6 (16 \D_B \D_{F_1} \D_{F_2}-(x-\D_{F_1}-\D_{F_2}-\D_B)^2)
 & 
x_3<x<x_4 \\
 96 \D_B \D_{F_1} \D_{F_2} 
 & 
 x>x_4\\
\end{array}
\right.
\end{split}
\end{equation}
}
where
$x_1 = \D_{F_1}+\D_{F_2} -\D_B$,
$x_2 =  \D_B+\D_{F_1}-\D_{F_2}$,
$x_3 = \D_B-\D_{F_1}+\D_{F_2} $,
and 
$x_4 =\D_B+\D_{F_1}+\D_{F_2}  $.    
    This time, we can observe that $f[x]$ is manifestly non-constant in a neighborhood of $x=0$, where a unique minimum is located (see figure \ref{ConifoldFx1}).
\begin{figure}[!h]
\centering
  \includegraphics[width=0.6\textwidth]{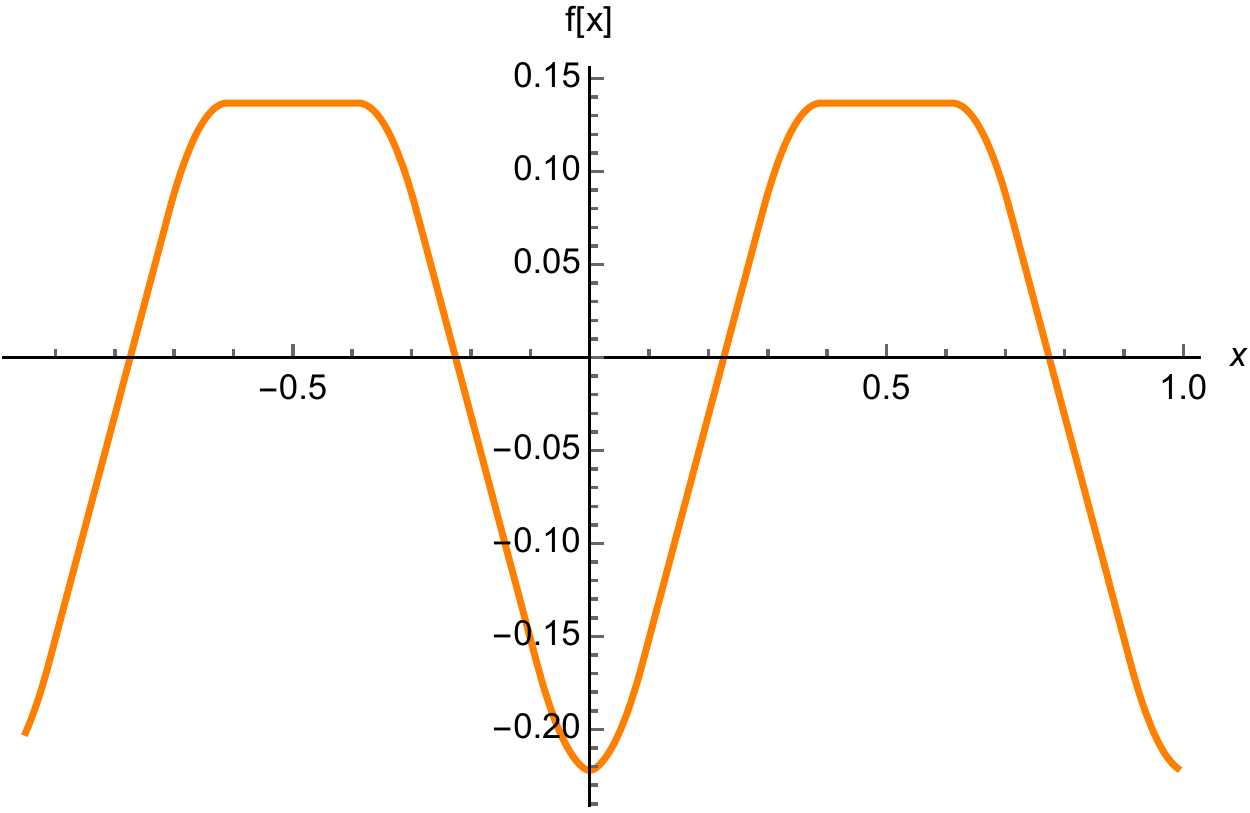}
    \caption{Plot of $f[x]$ for $\{\D{F_1},\D_{F_2},\D_{B}\}\,=\,\{0.05\,,0.15\,,0.19\}$}
    \label{ConifoldFx1}
    \end{figure}
Reminding that $V_2$ is actually, $-f[x]-f[y]$, we discover that this extremal point $a_+=a_-=0$, {\it i.e.} vanishing holonomies, dominates the Cardy-like limit in the regime:
    \begin{equation}
    \label{regimeB}
    \text{Re}\left(\frac{i}{\omega_1 \omega_2 }\right)>0\,.
    \end{equation}
    \end{itemize}
We showed that not all the chambers lead to honest isolated extrema and a similar analysis must be always performed. A numerical analysis for higher ranks still validates the work-hypothesis of vanishing holonomies and from now on we will consider arbitrary rank $N$ following this assumption\footnote{In appendix \ref{rank2Conifold} we perform a similar analysis for the conifold at rank 2; we show that it is reasonable to extend to this case the results of our rank-1 study}.

In order to exploit the geometric insight, from now on we prefer to take a suitable basis of field charges that are directly suggested by the toric diagram, as discussed in section \ref{Sec:Cardy}; in this basis flavour and baryonic symmetries get mixed and can be considered on equal footing. We will label the three $U(1)$ global symmetries simply as $U(1)_{1,2,3}$ and the associated fugacities as $\D_{1,2,3}$; $R$-symmetry will be denoted by $U(1)_R$ instead.
Following the discussion in section \ref{Sec:Cardy} we can parameterize the global charges from the geometry using perfect matchings.
\be 
\label{PMT11}
\scalebox{1}
{
\begin{tikzpicture}[font=\scriptsize, baseline] \begin{scope}[auto,%
  every node/.style={draw}, node distance=1cm]; 
\node[draw=none] at (-0.43, 0.57) {\tiny{2}};
\node[draw=none] at (-0.75, 1.4) {\tiny{1}};
\node[draw=none] at (-0.5, 2.15) {\tiny{2}};
\node[draw=none] at (0.15, 2.15) {\tiny{1}};
\node[draw=none] at (0.3, 1.4) {\tiny{2}};
\node[draw=none] at (1.2, 1.4) {\tiny{1}};
\node[draw=none] at (0.15, 0.5) {\tiny{1}};
\end{scope}

\draw[-o] (-0.3,0.3)--(-0.3,2.08);
\draw (-0.3,2.08)--(-0.3,2.5);
\draw (-0.7, 0.7)--(0.7,0.7);
\draw (-0.8, 2)--(-0.37,2);
\draw (-0.23,2)--(0.78,2);
\draw (1.1,1.1)--(1.1,1.71);

\draw[line width=0.4mm, blue] (-0.3, 0.7)--(0.7,0.7);
\draw[line width=0.4mm, blue](-0.8,2)--(-0.37,2);
\node[draw=none] at (-0.3,0.7) {$\bullet$};

\draw[thick,red] (0,0)--(-1.4,1.4);
\draw[thick,red] (0,0)--(1.4,1.4);
\draw[thick,red] (-1.4,1.4)--(0,2.8);
\draw[thick,red] (0,2.8)--(1.4,1.4);

\end{tikzpicture}} \qquad 
\scalebox{1}
{
\begin{tikzpicture}[font=\scriptsize, baseline] \begin{scope}[auto,%
  every node/.style={draw}, node distance=1cm]; 
\node[draw=none] at (-0.43, 0.57) {\tiny{2}};
\node[draw=none] at (-0.75, 1.4) {\tiny{1}};
\node[draw=none] at (-0.5, 2.15) {\tiny{2}};
\node[draw=none] at (0.15, 2.15) {\tiny{1}};
\node[draw=none] at (0.3, 1.4) {\tiny{2}};
\node[draw=none] at (1.2, 1.4) {\tiny{1}};
\node[draw=none] at (0.15, 0.5) {\tiny{1}};
\end{scope}

\draw[-o] (-0.3,0.3)--(-0.3,2.08);
\draw (-0.3,2.08)--(-0.3,2.5);
\draw (-0.7, 0.7)--(0.7,0.7);
\draw (-0.8, 2)--(-0.37,2);
\draw (-0.23,2)--(0.78,2);
\draw (1.1,1.1)--(1.1,1.71);

\node[draw=none] at (-0.3,0.7) {$\bullet$};
\draw[line width=0.4mm, blue] (-0.3, 0.76)--(-0.3,1.92);
\draw[thick,red] (0,0)--(-1.4,1.4);
\draw[thick,red] (0,0)--(1.4,1.4);
\draw[thick,red] (-1.4,1.4)--(0,2.8);
\draw[thick,red] (0,2.8)--(1.4,1.4);

\end{tikzpicture}}
 \qquad 
\scalebox{1}
{
\begin{tikzpicture}[font=\scriptsize, baseline] \begin{scope}[auto,%
  every node/.style={draw}, node distance=1cm]; 
\node[draw=none] at (-0.43, 0.57) {\tiny{2}};
\node[draw=none] at (-0.75, 1.4) {\tiny{1}};
\node[draw=none] at (-0.5, 2.15) {\tiny{2}};
\node[draw=none] at (0.15, 2.15) {\tiny{1}};
\node[draw=none] at (0.3, 1.4) {\tiny{2}};
\node[draw=none] at (1.2, 1.4) {\tiny{1}};
\node[draw=none] at (0.15, 0.5) {\tiny{1}};
\end{scope}

\draw[-o] (-0.3,0.3)--(-0.3,2.08);
\draw (-0.3,2.08)--(-0.3,2.5);
\draw (-0.7, 0.7)--(0.7,0.7);
\draw (-0.8, 2)--(-0.37,2);
\draw (-0.23,2)--(0.78,2);
\draw (1.1,1.1)--(1.1,1.71);

\node[draw=none] at (-0.3,0.7) {$\bullet$};
\draw[line width=0.4mm, blue] (-0.23, 2)--(0.8,2);
\draw[line width=0.4mm, blue] (-0.7, 0.7)--(-0.355,0.7);
\draw[thick,red] (0,0)--(-1.4,1.4);
\draw[thick,red] (0,0)--(1.4,1.4);
\draw[thick,red] (-1.4,1.4)--(0,2.8);
\draw[thick,red] (0,2.8)--(1.4,1.4);

\end{tikzpicture}}
 \qquad 
\scalebox{1}
{
\begin{tikzpicture}[font=\scriptsize, baseline] \begin{scope}[auto,%
  every node/.style={draw}, node distance=1cm]; 
\node[draw=none] at (-0.43, 0.57) {\tiny{2}};
\node[draw=none] at (-0.75, 1.4) {\tiny{1}};
\node[draw=none] at (-0.5, 2.15) {\tiny{2}};
\node[draw=none] at (0.15, 2.15) {\tiny{1}};
\node[draw=none] at (0.3, 1.4) {\tiny{2}};
\node[draw=none] at (1.2, 1.4) {\tiny{1}};
\node[draw=none] at (0.15, 0.5) {\tiny{1}};
\end{scope}

\draw[-o] (-0.3,0.3)--(-0.3,2.08);
\draw (-0.3,2.08)--(-0.3,2.5);
\draw (-0.7, 0.7)--(0.7,0.7);
\draw (-0.8, 2)--(-0.37,2);
\draw (-0.23,2)--(0.78,2);
\draw (1.1,1.1)--(1.1,1.71);

\node[draw=none] at (-0.3,0.7) {$\bullet$};
\draw[line width=0.4mm, blue] (-0.3,0.3)--(-0.3, 0.65);
\draw[line width=0.4mm, blue] (-0.3, 2.08)--(-0.3,2.5);
%\draw[line width=0.4mm, blue] (-0.3, 0.76)--(-0.3,1.92);
\draw[thick,red] (0,0)--(-1.4,1.4);
\draw[thick,red] (0,0)--(1.4,1.4);
\draw[thick,red] (-1.4,1.4)--(0,2.8);
\draw[thick,red] (0,2.8)--(1.4,1.4);

\end{tikzpicture}}
\ee
In this case there are four perfect matchings associated to the four external points of the toric diagram.  
They are listed in (\ref{PMT11}), where it is possible to observe that
in this case each PM corresponds to a bifundamental chiral field.
 The charges of these fields can be parameterized in terms
of the PMs as  
\begin{equation}
\begin{array}{c|c|c|c}
A_1\quad&  B_1\quad& A_2\quad&\quad B_2\quad\,\\
\hline
{\quad\bf a_1\quad}&{\quad\bf a_2\quad}&{\quad\bf a_3\quad}&{\quad\bf a_4\quad}
\end{array}
\end{equation}
The charges of the fields with respect to the symmetries suggested by the geometric data  
can be taken then as follows
\begin{equation}
\begin{tabular}{l | c c c c}
    &    $U(1)_1$    &    $U(1)_2$   &    $U(1)_3$ &    $U(1)_R$ \\
\hline
$A_1$   &   $1$   &   $0$   &  $0$  &   $1/2$\\
$A_2$     &   $0$   &   $0$   &  $1$   &   $1/2$\\
\hline
$B_1$    &   $0$   &   $1$   &   $0$  &   $1/2$\\
$B_2$    &   $-1$   &   $-1$  &  $-1$  &   $1/2$ \\
\end{tabular}
\end{equation}
We want to stress that, as in the $\mathcal{N}=4$ case discussed in section \ref{Sec:Cardy},   the $R$-symmetry
that we consider here accidentally coincides with the exact $R$-symmetry at the conformal fixed point.
However this will not be the case in the models that we are going to discuss in the next section.
Observe that the new 
conjugated fugacities $\Delta_i$ can thought as linear combination of the flavour and baryonic ones:
\begin{equation}
\D_1\,=\, \D_B+\D_{F_1}+\D_{F_2}\,,\quad \D_2\,=\,-\D_B+\D_{F_1}-\D_{F_2}\,,\quad \D_3\,=\, \D_B-\D_{F_1}-\D_{F_2}\,.
\end{equation}
$V_2$ for arbitrary rank can be now expressed as:
\begin{equation}
\begin{split}
V_2\,=\,-\sum_{m, n=1}^N &\left( \kappa\left[ a^{(1)}_m-a^{(2)}_n+\D_1 \right] + \kappa\left[ a^{(2)}_m-a^{(1)}_n+\D_2 \right] +\right.\\
&\left.\,\,\,\,\kappa\left[ a^{(1)}_m-a^{(2)}_n+\D_3 \right] + \kappa\left[ a^{(2)}_m-a^{(1)}_n-\D_1-\D_2-\D_3 \right] \right)\,.
\end{split}
\end{equation}
In this basis we  fix the fugacities such that $0\leq \D_{1,2,3}<1/2$ and $0\leq \D_1+\D_2+\D_3 \leq 1/2$.
In this chamber, $f[x]$ enjoys a local maximum for vanishing holonomies and thus $V_2$ exhibits a minimum\footnote{The range we fixed for $\Delta_{1,2,3}$ leads to slightly different features with respect to the one we chose for $\D_{F_{1,2}},\D_B$ in our previous discussion; in the former case, $V_2$ enjoys a local minimum while in the latter $V_2$  possesses a local maximum, in both cases for vanishing holonomies. For this reason, the two extrema dominates the Cardy-like limit in different regimes, \eqref{regimeX} and $\eqref{regimeB}$ respectively. Both fugacity ranges can be chosen, up to minimal changes to be performed in going from one regime to the other, as carefully shown in \cite{ArabiArdehali:2019tdm}.}. 
The extremum dominates the Cardy-like limit if
 \begin{equation}
 \label{regimeX}
    \text{Re}\left(\frac{i}{\omega_1 \omega_2}\right)<0\,.
 \end{equation}

In the fixed regime, we can evaluate the dominant saddle contribution:
\begin{equation}
\begin{split}
V\,=&\,\frac{i\,\pi}{2\omega_1 \omega_2}\left( \frac{V_2}{3}+(\omega_1+ \omega_2) V_1\right)\bigg|_{a^{(I)}_m=0}\,=\,\\
=&\frac{\pi i\,N^2}{2\omega_1 \omega_2}\left\{ (\omega_1+ \omega_2)\left(\D_3(\D_3-1)+\D_1(\D_2+\D_3-1)+\D_1^2+\D_2^2+\D_2(\D_3-1)\right) +\right.\\
&\left.-\left(\D_2\D_3(\D_2+\D_3-1)+\D_1^2(\D_2+\D_3+\D_1(\D_2+\D_3-1)(\D_2+\D_3))\right)\right\}\,.
 \end{split}
\end{equation}
This equation get further simplified performing a suitable shift of the fugacities
\begin{equation}
\label{shiftconifold}
\D_{1,2,3}\,\rightarrow\, \D_{1,2,3}-\frac{\omega_1 + \omega_2}{4}
\end{equation}
and taking the leading order in the Cardy-like limit $|\omega_1|, |\omega_2|\rightarrow 0\,\,$;
let us stress that the shift \eqref{shiftconifold} is actually dictated by the geometry: as we will test for other toric models in the next sections, the fugacities get always shifted by $-\frac{\omega_1 + \omega_2}{d}$ where $d$ is the number of external points in the toric diagram.
After the shift, the entropy function can be expressed as
\begin{equation}
\label{entropyConifold}
S_E\,=\,-\frac{ i \pi N^2}{\omega_1\omega_2}(\D_1\D_2\D_3+\D_1\D_2\D_4+\D_1\D_3\D_4+\D_2\D_3\D_4)\,,
\end{equation}
where we have defined:
\begin{equation}
\D_4\,\equiv \,\omega_1+\omega_2+1-\D_1-\D_2-\D_3\,.
\end{equation}
The entropy function \eqref{entropyConifold}  enjoys the expected scaling behaviour $S_E\,\propto N^2$; moreover, it is in perfect agreement with our general proposal
\begin{equation}
S_E\,=\,-\frac{i \pi N^2}{6\,\omega_1 \omega_2}C_{IJK}\D_I\D_J\D_K\,,
\end{equation}
 where $I,J,K$ run from $1$ to $4$ and $C_{IJK}$ is defined as in \eqref{BPT}.

\section{Other examples}
\label{Sec:OtherExamples}
In this section we test our proposal \eqref{expected} in various cases of growing complexity. 
In each case we assign the charges using the prescription discussed in section \ref{Sec:Cardy}
and we assume that the Cardy-like limit is dominated by a unique minimum where all the holonomies vanish.
We have tested the last conjecture in the rank-1 cases of $\text{dP}_1$,$\text{dP}_2, \text{(P)dP}_4$ and $\mathbb{F}_0$, 
finding evidence of its validity. In each case the minimum is found in a chamber where fugacities 
$\D_i$ of the $d-1$ U(1) global symmetries are taken such that:
\be
0\leq \D_i\leq \frac{1}{2}\,\,\,\forall i\,,\quad 0\leq \sum_{i=1}^{d-1}\,\D_i\,\leq 1\,.
\ee
Since in this range $V_2$ enjoys a minimum, we restrict to the regime \eqref{regimeX}. 

As a general remark  let us stress that some of the theories that we are going to discuss admit more Seiberg dual realizations, 
denoted as phases. We specify for each model the Seiberg phase that we focus on.
Finally, observe that we are not necessarily fixing the $R$-charges of the fields at the conformal fixed point, 
but we refer to a trial $R$-current, using the uniform prescription for all the models under investigation,
as explained in section \ref{Sec:Cardy}.

\subsection{SPP}
\label{SPP}
The suspended pinch point (SPP) gauge theory 
corresponds to the near horizon limit of a stack of $N$  D3 branes
probing  the tip of the conical singularity,  $x^2 y = w z$.
This is the simplest example of a larger class of models, defined by the equation
$x^a y^b = w z$, denoted as $L^{aba}$ models.
In the SPP case the toric Sasaki-Einstein base in described by the following vectors:
\begin{equation}
	V_1=(1,1,1), \quad V_2=(1,0,1),\quad V_3=(0,0,1), \quad V_4=(2,0,1),\quad V_5=(1,0,1)\,. 
\end{equation}
The vector $V_5$ represents a point on the perimeter of the toric diagram and it turns out 
that two different perfect matchings can be associated to it and, consequently, we can get 
two different possible charge sets.
Following the prescription of \cite{Butti:2005vn}  we can associate a non vanishing set of charges
to just one of them.

The theory living on a stack of $N$ D3-branes at the SPP conical singularity is described by the following quiver:
\be 
\scalebox{1.2}
{
\begin{tikzpicture}[font=\scriptsize, baseline] \begin{scope}[auto,%
  every node/.style={draw}, node distance=1cm]; 
 \node[draw=none] at (0,0) {$\bullet$};
  \node[draw=none] at (0,-0.3) {$3$};
 \node[draw=none] at (3,0) {$\bullet$};
 \node[draw=none] at (3.,-0.3) {$2$};
 \node[draw=none] (pinch) at (1.5,2.3) {$\bullet$};
  \node[draw=none] at (1.5,2.) {$1$};
   \node[draw=none] at (0.1,0.7) {$X_{13}$};
      \node[draw=none] at (0.7,1.8) {$X_{31}$};
       \node[draw=none] at (3.,0.7) {$X_{12}$};
      \node[draw=none] at (2.3,1.8) {$X_{21}$};
          \node[draw=none] at (0.9,-0.3) {$X_{23}$};
      \node[draw=none] at (2.1,-0.3) {$X_{32}$};
 %\node[draw=none] at (2.4, 1) {$X^{(\alpha)}_{23}$};
% \node[draw=none] at (1, 2.4) {$X_{12}$};
 \end{scope}
 %\draw[black] () edge[out=135,in=45, loop, looseness=8] node[midway,above=0.1cm] {$\phi$} (pinch);
 \draw[black] (1.5,2.3) edge[out=150,in=-60] (1.2,2.6);
  \draw[black] (1.2,2.6) edge[out=120,in=180] node[midway,above=0.05cm] {$\phi$} (1.5,3);
    \draw[black] (1.5,3) edge[out=0,in=60]  (1.8,2.6);
 \draw[black] (1.8,2.6) edge[out=240,in=30] (1.5,2.3);
 \draw[black, bif](0,0)--(3,0);
  \draw[black, bif](0,0)--(1.5,2.3);
   \draw[black, bif](1.5,2.3)--(3,0);
 \end{tikzpicture}}
\ee
with superpotential
\be
W=\Tr[X_{21}X_{12}X_{23}X_{32}-X_{32}X_{23}X_{31}X_{13}+X_{13}X_{31}\phi-X_{12}X_{21}\phi].
\ee
Each $X_{ij}$ transforms in the $\mathbf{N}$ representation of the $i$ node and in the $\overline{\mathbf{N}}$ of the $j$-th node; the field transforming in the adjoint of the first node is named, instead, $\phi$.
The charges of the fields can be parameterized in terms of the PMs using the assignation
\begin{equation}
\label{SPPPM}
\begin{array}{c|c|c|c|c|c|c}
\phi \quad&X_{12}\quad& X_{21}\quad&X_{23}\quad&X_{32}\quad& X_{31}\quad& X_{13}\quad\\
\hline
\mathbf{\quad a_1+a_2\quad }& \mathbf{\quad a_4 \quad}&\mathbf{\quad a_3+a_5 \quad}& \mathbf{\quad a_2 \quad} & \mathbf{ \quad a_1 \quad} & \mathbf{ \quad a_4+a_5 \quad} &\mathbf{\quad a_3 \quad}
\end{array}
\end{equation}
It follows that the charge assignment for $U(1)_R$ and the extra four $U(1)_i$  global can be taken as follows:
\begin{equation}
\begin{tabular}{c|c c c c c}
 &    $U(1)_1$    &    $U(1)_2$   &    $U(1)_3$ & $U(1)_4$ &    $U(1)_R$ \\
\hline
$\phi$  & 1 & 1 & 0 & 0 & 4/5\\
\hline
$X_{12}$  & 0 & 0 & 0 & 1 & 2/5\\
$X_{21}$ & -1 & -1 & 0 & -1 & 4/5\\
\hline 
$X_{23}$  & 0 & 1 & 0 & 0 & 2/5\\
$X_{32}$  & 1 & 0 & 0 & 0 & 2/5\\
\hline
$X_{31}$  & -1 & -1 & -1 & 0 & 4/5\\
$X_{13}$  & 0 & 0 & 1 & 0 & 2/5\\
\end{tabular}
\end{equation}
Observe that as usual we took a combination of $U(1)$ natural for toric geometry and we have not done any distinction between flavour and baryonic symmetries. We denote
$\D_i$ the fugacity associated to $U(1)_i$. With this assignment, $V_2$ admits a minimum for vanishing holonomies in a chamber where $0\leq \D_i\leq 1/2$ for each $U(1)_i$ and $0< \D_1+\D_2+\D_3+\D_4<1$; we can evaluate 
\be
	V=\frac{i \pi}{2 \omega_1 \omega_2}\left(V_1 (\omega_1+\omega_2)+\frac{V_2}{3}\right)
\ee
and, after performing a shift of the charges by a factor $-\tfrac{\omega_1+\omega_2}{5}$ 
and taking the leading order in $|\omega_1|,|\omega_2|\rightarrow 0$, we get:
\begin{equation}
\label{VleadSPP}
\begin{split}
	V_{\text{leading}}\,=\,\frac{i\pi N^2}{\omega_1\omega_2}&\left(\D_1((\D_2+\D_3)(\D_2+\D_3\!-\!\omega_1\!-\!\omega_2\!-\!1)+
	\D_4^2\!-\!\D_4(1\!+\!\omega_1\!+\!\omega_2\!-\!2\D_2))\right.\\
	&\left.\,\,\,\,+ \D_2(\D_3^2-\D_3(1+\omega_1+\omega-\D_2)+\D_4(\D_2+\D_4-1-\omega_1-\omega_2))+\right.\\
	&\,\,\,\,\left.+\D_1^2(\D_2+\D_3+\D_4)\right).
\end{split}
\end{equation}
If we now define a new constrained fugacity:
\be
\quad\D_5\,=\,1+\omega_1+\omega-\D_1-\D_2-\D_3-\D_4
\ee
we obtain the following expression for the entropy:
\be
\begin{split}
	S_{E}=-\frac{i\pi N^2}{\omega_1\omega_2}&\left( \D_1 \D_2 \D_3+\D_1\D_2 \D_4 +2 \D_1 \D_3 \D_4+2 \D_2\D_3\D_4+\right.\\ 
	&\,\,\,\,\left.+\D_1 \D_2 \D_5+\D_1\D_3\D_5+\D_2\D_3\D_5+\D_1 \D_4 \D_5+\D_2\D_4\D_5\right)\,.
	\end{split}
\ee 
This result is in agreement with our expectation from toric geometry,  encoded in formula \eqref{expected}. 

Finally, as discussed at the beginning of this section, the SPP singularity  can be thought as a particular case of a larger 
class of toric models, denoted as $L^{aba}$, for $a=1\,,b=2$.
The toric diagram of an $L^{aba}$ singularity is depicted in (\ref{ciccio}).
\be 
\label{ciccio}
\scalebox{0.8}
{
\begin{tikzpicture}[font=\scriptsize, baseline] \begin{scope}[auto,%
  every node/.style={draw}, node distance=1cm]; 
 \node[draw=none] at (0,-1) {$\bullet$};
 \node[draw=none] at (0,1) {$\bullet$};
  \node[draw=none] at (2,-1) {$\bullet$};
    \node[draw=none] at (4,-1) {$\bullet$};
        \node[draw=none] at (2,1) {$\bullet$};
                \node[draw=none] at (0,1) {$\bullet$};
 %\node[draw=none] at (2.4, 1) {$X^{(\alpha)}_{23}$};
% \node[draw=none] at (1, 2.4) {$X_{12}$};
 \end{scope}
 \draw[black] (0,-1)--(2,-1)--(4,-1)--(2,1)--(0,1)--(0,-1);
 \end{tikzpicture}}
 \quad\Rightarrow\quad
 \scalebox{0.8}
{
\begin{tikzpicture}[font=\scriptsize, baseline] \begin{scope}[auto,%
  every node/.style={draw}, node distance=1cm]; 
 \node[draw=none] at (0,-1) {$\bullet$};
 \node[draw=none] at (0,1) {$\bullet$};
  \node[draw=none] at (2,1) {$\bullet$};
    \node[draw=none] at (2,-1) {$\bullet$};
      \node[draw=none] at (6,-1) {$\bullet$};
   \node[draw=none] at (2,1.5) {$a-1$ points on the perimeter};
      \node[draw=none] at (3,-1.5) {$b-1$ points on the perimeter};
        \node[draw=none] at (4,1) {$\bullet$};
                \node[draw=none] at (0,1) {$\bullet$};
 %\node[draw=none] at (2.4, 1) {$X^{(\alpha)}_{23}$};
% \node[draw=none] at (1, 2.4) {$X_{12}$};
 \end{scope}
 \draw[black] (0,-1)--(2,-1);
 \draw[black, thick, dotted] (2.7,-1)--(5.3,-1);
 \draw[black] (0,-1)--(0,1)--(2,1);
  \draw[black,thick, dotted] (2.4,1)--(3.6,1);
    \draw[black] (2,-1)--(2.7,-1);
        \draw[black] (2,1)--(2.4,1);
                \draw[black] (4,1)--(3.6,1);
        \draw[black] (6,-1)--(5.3,-1);
   \draw[black] (6,-1)--(4,1);
 \end{tikzpicture}}
\ee
In this case there are $a+b$ gauge groups, and two flavor symmetries and $a+b-1$ non anomalous baryonic symmetries.
This huge amount of baryonic symmetries reflects in the toric diagram onto the large number of external point 
lying on the perimeter.
Each of these points contribute with triangle areas to reproduce the correct entropy function, 
following the general prescription \eqref{expected}. 
Observe that for $a=0$ the models become $\mathcal{N}=2$ necklace quivers, corresponding to 
$\mathbb{Z}_b$ orbifolds of $\mathcal{N}=4$ SYM.
 The entropy for these models has been studied in \cite{Honda:2019cio}, by turning off the baryonic fugacities.
Here in section \ref{Sec:Legendre}  we will study the most general situation.

\subsection{$\mathbb F_0$}
The complex cone over the first Hirzebruch surface $\mathbb F_0$ is a $\mathbb Z_2$ orbifold of the conifold; the toric diagram is parametrized by the four vectors
\begin{equation}
	V_1=(0,0,1), \quad V_2=(1,0,1),\quad V_3=(0,2,1), \quad V_4=(-1,2,1). 
\end{equation}
The corresponding theory in its phase $I$ is described by the following quiver and superpotential: 

\begin{equation}
\scalebox{1.2}
{
\begin{tikzpicture}[font=\scriptsize, baseline] \begin{scope}[auto,%
  every node/.style={draw}, node distance=1cm]; 
 \node[draw=none] at (0,-1) {$\bullet$};
 \node[draw=none] at (0,1) {$\bullet$};
 \node[draw=none] at (2,-1) {$\bullet$};
 \node[draw=none] at (2,1) {$\bullet$};
 \node[draw=none] at (1, -1.4) {$X^{(\a)}_{34}$};
 \node[draw=none] at (-0.4, 0) {$X^{(\a)}_{41}$};
 \node[draw=none] at (2.5, 0) {$X^{(\a)}_{23}$};
 \node[draw=none] at (1, 1.4) {$X^{(\a)}_{12}$};
  \node[draw=none] at (-0.2,1.2) {1};
   \node[draw=none] at (2.2,1.2) {2};
    \node[draw=none] at (2.2,-1.2) {3};
 \node[draw=none] at (-0.2, -1.2) {4};
 \end{scope}
 \draw (0,-1)--(0,1);
 \draw[->>] (0,-1)--(0,0);
 \draw[<<-] (0.9,-1)--(2,-1);
 \draw (0,-1)--(2,-1);
 \draw (2,-1)--(2,1);
 \draw (0,1)--(2,1);
 \draw[->>] (2,1)--(2,-0.1);
 \draw[->>] (0,1)--(1,1);
 \end{tikzpicture}}
\quad\quad
	W=\epsilon_{\a\b}\epsilon_{\g\delta}\text{Tr}[X^{(\a)}_{12}X^{(\b)}_{34}X^{(\g)}_{23}X^{(\delta)}_{41}].
\end{equation}
One can assign charges to the fields in the theory directly from the geometry. 
The charges of the fields can be parameterized in terms of the PMs using the assignation
\begin{equation}
\begin{array}{c|c|c|c}
X_{12} &X_{34} &X_{23} &X_{41}\\
\hline
{\bf a_1}&{\bf a_2}&{\bf a_3}&{\bf a_4}
\end{array}
\end{equation}
The model has three global $U(1)$ symmetries in addition to one $U(1)_R$ and in this case one gets the following global charges
\begin{equation}
\label{tableF}
\begin{array}{c|c|cccc}
  &\text{multiplicity} &\,U(1)_1 & U(1)_2 & U(1)_3 &U(1)_R\\
\hline
X_{12}     &2&  1&0&0 &1/2 \\ 
X_{23}    &2& 0&1&0  &1/2 \\
X_{34}     &2& 0 &0 &1 & 1/2 \\
X_{41}     &2& -1 &-1 &-1& 1/2 \\ 
\end{array}
\end{equation}
We now compute the Cardy-like limit of the superconformal index for this theory; if we denote the fugacities for the symmetries $U(1)_{1,2,3}$ as $\D_{1,2,3}$ respectively, the expression that we find after shifting each fugacity by $-\tfrac{\o_1+\o_2}{4}$ is 
\be\label{VF0}
\begin{split}
	V_{\text{leading}}=-\frac{2\pi i N^2}{\o_1\o_2} \bigl(&\D_2(1+\o_1 +\o_2-\D_2-\D_3)\D_3-\D_1^2(\D_2+\D_3)\\
	&\bigl.-\D_1(\D_2+\D_3)(1+\o_1+\o_2-\D_2-\D_3)\bigr)
\end{split}
\ee
The entropy function in this case can be written as:
\be
	S_E(\D,\o)=-\frac{2 \pi iN^2}{\o_1\o_2}(\Delta_1 \Delta_2 \Delta_3+\Delta_1 \Delta_2 \Delta_4+\Delta_1 \Delta_3 \Delta_4+	\Delta_2 \Delta_3 \Delta_4),
\ee
and it exactly reproduces \eqref{VF0} by using the constraint
\be
	\Delta_4=1+\o_1 +\o_2-\D_1-\D_2-\D_3,
\ee
Observe that the entropy function just obtained is twice the one for the conifold, as one should expect from the fact that we are dealing with a $\mathbb Z_2$ orbifold of the latter
\footnote{To be more precise, the entropy function reproduces twice the conifold one because the orbifold action does not introduce new singularities or, equivalently,  
new symmetries. A non-chiral $\mathbb{Z}_2$ orbifold of the conifold like the $L^{222}$ model does not have this 
property.}.

\subsection{dP$_1$}
Let us consider the theory arising from a stack of $N$ D3 branes at the tip of the complex Calabi-Yau cone whose base is the first del Pezzo surface. The toric diagram is generated by 
\be
	V_1=(1,0,1), \quad V_2=(0,1,1),\quad V_3=(-1,0,1), \quad V_4=(-1,-1,1). 
\ee
The corresponding quiver is as follows 
\be 
\scalebox{1.2}
{
\begin{tikzpicture}[font=\scriptsize, baseline] \begin{scope}[auto,%
  every node/.style={draw}, node distance=1cm]; 
 \node[draw=none] at (0,0) {$\bullet$};
 \node[draw=none] at (0,2) {$\bullet$};
 \node[draw=none] at (2,0) {$\bullet$};
 \node[draw=none] at (2,2) {$\bullet$};
 \node[draw=none] at (1, -0.3) {$X^{(\a)}_{34}$, $X^{(3)}_{34}$};
 \node[draw=none] at (-0.4, 1) {$X^{(\a)}_{41}$};
 \node[draw=none] at (2.4, 1) {$X^{(\a)}_{23}$};
 \node[draw=none] at (1, 2.4) {$X_{12}$};
 \node[draw=none] at (1.2, 0.4) {$X_{13}$};
 \node[draw=none] at (1.2, 1.6) {$X_{42}$};
  \node[draw=none] at (-0.2,2.2) {1};
   \node[draw=none] at (2.2,2.2) {2};
    \node[draw=none] at (2.2,-0.2) {3};
 \node[draw=none] at (-0.2, -0.2) {4};
 \end{scope}
 \draw (0,0)--(0,2);
 \draw[->>] (0,0)--(0,1);
 \draw[<<<-] (0.9,0)--(2,0);
 \draw (0,0)--(2,0);
 \draw (2,0)--(2,2);
 \draw (0,2)--(2,2);
 \draw[->>] (2,2)--(2,0.9);
 \draw[->] (0,2)--(1,2);
 \draw (0,0)--(2,2);
 \draw[->] (0,0)--(1.414,1.414); 
 \draw[->] (0,2)--(1.414,0.590); 
 \draw (0,2)--(2,0);
 \end{tikzpicture}}
\ee
and the superpotential for this theory reads
\be
	W=\epsilon_{\a\b}\Tr	[X^{(\a)}_{34}X^{(\b)}_{41}X_{13}-X^{(\a)}_{34}X^{(\b)}_{23}X_{42}+
	X_{12}X^{(3)}_{34}X^{(\a)}_{41}X^{(\b)}_{23}].
\ee
The charges of the fields can be parameterized in terms of the PMs using the assignation
\begin{equation}
\begin{array}{c|c|c|c|c|c}
X_{34}^{(3)},X_{13},X_{42} &X_{23}^{(1)},X_{41}^{(1)} & X_{12} &X_{23}^{(2)},X_{41}^{(2)} &X_{34}^{(1)} &X_{34}^{(2)} \\
\hline
{\bf a_1}&{\bf a_2}&{\bf a_3}&{\bf a_4}&{\bf a_2+a_3}&{\bf a_3+a_4}
\end{array}
\end{equation}
The charges for $U(1)_R$ and the three global $U(1)$ symmetries of the model coming from the perfect matching are the following 
\begin{equation}
\label{tableF}
\begin{array}{c|cccc}
&U(1)_1 & U(1)_2 & U(1)_3 &U(1)_R\\
\hline
X_{12}     	      &0& 0&1 & 1/2 \\ 
X_{23}^{(1)}    &0& 1&0  &1/2 \\
X_{23}^{(2)}    &-1& -1&-1 &1/2 \\
\hline
X_{34}^{(1)}    &0& 1 &1 &1\\
X_{34}^{(2)}    &-1& -1 &0 & 1 \\
X_{34}^{(3)}    &1& 0 &0 &1/2\\
\hline
X_{41}^{(1)}    &0& 1 &0 &1/2 \\ 
X_{41}^{(2)}    &-1& -1 &-1 &1/2 \\ 
\hline
X_{13}	      &1& 0 &0&1/2 \\ 
X_{42}	      &1&0 &0 &1/2 \\ 
\end{array}
\end{equation}
The expression for the entropy function in this case gives:
\be\label{SEdP1}
	S_{E}(\D,\o)=-\frac{i\pi N^2}{\o_1\o_2} (2\D_1\D_2\D_3+2\D_1\D_2\D_4+2\D_1\D_3\D_4+\D_2\D_3\D_4).
\ee 
Again, the same result can be obtained by taking the Cardy-like limit of the superconformal index. The leading order of the function
\be
	V=\frac{i \pi N^2}{2 \o_1 \o_1}\left(V_1 (\o_1+\o_2)+\frac{V_2}{3}\right),
\ee
taken after shifting the charges by a factor $-\tfrac{\o_1+\o_2}{4}$,
is given by
\be
\begin{split}\label{VleaddP1}
	V_{\text{leading}}=\frac{i\pi N^2}{\o_1\o_2} &\bigl(3 \D_1 \D_2(\D_1+\D_2-\o_1-\o_2-1)-2\D_1\D_3(1+\o_1+\o_2-\D_1)\bigr. \notag \\
&-(1+\o_1+\o_2-4\D_1)\D_2\D_3+\D_2^2\D_3+(2\D_1+\D_2)\D_3^2\bigr).
\end{split}
\ee
If we now take the expression of the entropy function \eqref{SEdP1} and impose the constraint on the fugacities
\be
	\Delta_4=1+\o_1 +\o_2-\D_1-\D_2-\D_3,
\ee
we obtain the expression \eqref{VleaddP1} for the entropy function.

\subsection{dP$_2$}
The toric diagram for the complex cone over the dP$_2$ surface is generated by the following vectors
\be
	V_1=(1,1,1), \quad V_2=(0,1,1),\quad V_3=(-1,0,1), \quad V_4=(-1,-1,1), \quad V_5=(0,-1,1). 
\ee
The charges of the fields can be parameterized in terms of the PMs using the assignation
\begin{equation}
\begin{array}{c|c|c|c|c|c|c|c|c|c|c}
X_{13} & X_{24} & X_{51}^{(1)} & X_{23} & X_{41} &X_{51}^{(2)} & X_{12}^{(2)} & X_{45} & X_{12}^{(1)} & X_{35} & X_{34} \\
\hline
{\bf a_4+a_5}&{\bf a_5}&{\bf a_5}&{\bf a_2}&{\bf a_1+a_2}&{\bf a_2+a_3}&{\bf a_3+a_4}&{\bf a_4}&{\bf a_1}&{\bf a_1}&{\bf a_3}
\end{array}
\end{equation}
The theory arising from a stack of $N$ D3 branes put at the tip of this toric Calabi-Yau cone admits two phases. The phase $I$ can be described by a quiver with five nodes
\be 
\scalebox{1.2}
{
\begin{tikzpicture}[font=\scriptsize, baseline] \begin{scope}[auto,%
  every node/.style={draw}, node distance=1cm]; 
 \node[draw=none] at (0,0) {$\bullet$};
 \node[draw=none] at (1.5,0) {$\bullet$};
 \node[draw=none] at (-0.3,1.2) {$\bullet$};
  \node[draw=none] at (1.8,1.2) {$\bullet$}; 
  \node[draw=none] at (0.75,2) {$\bullet$};
 \node[draw=none] at (0.75,2.2) {1};
 \node[draw=none] at (-0.5,1.2) {5};
   \node[draw=none] at (2,1.2) {2};
    \node[draw=none] at (1.75,-0.2) {3};
 \node[draw=none] at (-0.2, -0.2) {4};
 \end{scope}
 \draw[ddirected](-0.3,1.2)--(0.75,2);
  %\draw[->>](0.75,2.05)--(1.76,1.25);
  \draw[sdirected]  (1.5,0) --(0.05,0);
  \draw[sdirected]  (0,0)--(-0.3,1.15);
  \draw[sdirected]   (1.8,1.2)--(1.5,0.05);
   \draw[sdirected]  (1.5,0)-- (-0.25,1.2);
   \draw[sdirected]  (0,0)--(1.75,1.2);
 \draw[ddirected] (0.75,2)--(1.8,1.2);
     \draw[sdirected]  (0,0)--(0.75,2);
      \draw[sdirected]  (1.5,0)--(0.76,1.95);
      % \draw[directed] (0,0) -- (3,3);
       %\draw[directed](0.75,2.05)--(1.76,1.25);
 \end{tikzpicture}}
\ee
and superpotential 
\be	
\begin{split} 
	W=\Tr\bigl[&X_{13}X_{34}X_{41}-X_{12}^{(2)}X_{24}X_{41}+X_{12}^{(1)}X_{24}X_{45}X_{51}^{(2)}-X_{13}X_{35}X_{51}^{(2)}\bigr.\\
	&\bigl.+X_{12}^{(2)}X_{23}X_{35}X_{51}^{(1)}-X_{12}^{(1)}X_{23}X_{34}X_{45}X_{51}^{(1)}\bigr].
\end{split}
\ee
Here $X_{ij}$ denotes a bifundamental field connecting the $i$-th and $j$-th nodes. The charges assigned to the various fields in the quiver directly from the perfect matchings are 
\begin{equation}
\label{tableF}
\begin{array}{c|ccccc}
  &U(1)_1 & U(1)_2 & U(1)_3& U(1)_4 &U(1)_R\\
\hline
X_{13}     	      &-1& -1&-1 & 0&4/5 \\ 
X_{24}    	&-1& -1&-1 &-1&2/5 \\
X_{51}^{(1)}    &-1& -1&-1 &-1&2/5 \\
X_{23}    &0& 1 &0 &0&2/5\\
X_{41}    &1& 1 &0 &0&4/5\\
X_{51}^{(2)}     &0& 1 &1 &0&4/5\\
X_{12}^{(2)}     &0& 0 &1&1&4/5\\
X_{45}    &0& 0 &0 &1&2/5\\
X_{12}^{(1)}	     &1& 0 &0 &0&2/5\\
X_{35}	      &1& 0 &0 &0&2/5\\
X_{34}	     &0& 0 &1 &0&2/5\\
\end{array}
\end{equation}
The entropy function obtained from toric geometry is:
\be\label{SEdP2}
\begin{split}
	S_{E}(\D,\o)=-\frac{i\pi N^2}{\o_1\o_2}\bigl(&\D_1\D_2\D_3+2\D_1\D_2\D_4+2\D_1\D_3\D_4+			\D_2\D_3\D_4\bigr.\\
	&\bigl.+2\D_1\D_2\D_5+3\D_1\D_3\D_5+3\D_1\D_3\D_5+2\D_2\D_3\D_5\bigr.\\
	&+\bigl.2\D_1\D_4\D_5+2\D_2\D_4\D_5+\D_3\D_4\D_5\bigr)
\end{split}
\ee
The leading order of the Cardy-like limit of the superconformal index gives, after shifting the charges by a factor $-\tfrac{\o_1+\o_2}{5}$
\be
\begin{split}
	V_{\text{leading}}=\frac{i\pi}{\o_1\o_2}&\Bigl[ (2\D_2^2(\D_3+\D_4)+(\D_3\D_4+2\D_2(\D_3+\D_4))
	(\D_3+\D_4\!-\!1\!-\!\o_1\!-\!\o_2)\Bigr.	+\\
	\Bigl.&+\D_1^2(2\D_2+3\D_3+2\D_4)+2\D_1\D_2^2-3\D_1\D_3(1+\o_1+\o_2-\D_3)+\Bigr.\\
	&\Bigl.-2\D_1\D_2(1+\o_1+\o_1-3\D_3-2\D_4)-2\D_1\D_4(1+\o_1+\o_2-2\D_3)\Bigr.\\
	&\Bigl.+2\D_1\D_4^2\Bigr].
\end{split}
\ee
This is the same expression that one gets by taking \eqref{SEdP2} and using the constraint on the fugacities
\be
	\D_5=-\D_1-\D_2-\D_3-\D_4+\o_1+\o_2+1.
\ee

\subsection{dP$_3$}
The toric diagram for the Calabi-Yau cone over the dP$_3$ surface is generated by the following vectors
\be
\begin{split}
	&V_1=(1,1,1), \quad \quad \; \;V_2=(0,1,1),\quad \; \; \; V_3=(-1,0,1), \\
	&V_4=(-1,-1,1), \quad V_5=(0,-1,1), \quad V_6=(1,0,1). 
\end{split}
\ee
The charges of the fields can be parameterized in terms of the PMs using the assignation
\begin{equation}
\begin{array}{c|c|c|c|c|c|c|c|c|c|c|c}
X_{12} & X_{13} & X_{23} & X_{24} & X_{34} &X_{35} & X_{45} & X_{46} & X_{56} & X_{51} & X_{61} & X_{62} \\
\hline
{\bf a_6}&{\bf a_2+a_3}&{\bf a_5}&{\bf a_1+a_2}&{\bf a_4}&{\bf a_1+a_6}&{\bf a_3}&{\bf a_5+a_6}&{\bf a_2}&{\bf a_4+a_5}&{\bf a_1}&{\bf a_3+a_4}
\end{array}
\end{equation}

The theory associated to this cone admits three phases; in its phase $I$, it can be described by the following quiver:
\be 
\scalebox{1.2}
{
\begin{tikzpicture}[font=\scriptsize, baseline] \begin{scope}[auto,%
  every node/.style={draw}, node distance=1cm]; 
 \node[draw=none] (5) at (0,0)  {$\bullet$};
 \node[draw=none] (4) at  (1.2,0){$\bullet$};
 \node[draw=none] (6) at (-0.5,1)  {$\bullet$};
  \node[draw=none] (3) at (1.7,1) {$\bullet$};
  \node[draw=none] (1)  at (0,2) {$\bullet$};
  \node[draw=none] (2)  at (1.2,2)   {$\bullet$};
  
 \node[draw=none] at (-0.15,2.2) {1};
  \node[draw=none] at (1.35,2.2) {2};
 \node[draw=none] at (-0.7,1) {6};
  \node[draw=none] at (1.9,1) {3};
    \node[draw=none] at (1.35,-0.2) {4};
 \node[draw=none] at (-0.15, -0.2) {5};
 \end{scope}
 \draw[sdirected]  (0,2)--(1.2,2);
  \draw[sdirected]  (1.2,2)--(1.7,1);
  \draw[sdirected]  (1.7,1)--(1.2,0);
    \draw[sdirected] (1.2,0)--(0,0);
      \draw[sdirected] (0,0)--(-0.5,1);
   \draw[sdirected] (-0.5,1)--(0,2);
   \draw[sdirected] (0,0)-- (0,2); 
    \draw[sdirected] (1.2,2)--(1.2,0);  
    \draw[sdirected] (1.2,0)-- (-0.5,1);  
     \draw[sdirected] (1.7,1)-- (0,0);   
       \draw[sdirected] (-0.5,1)-- (1.2,2);
      \draw[sdirected]  (0,2)--(1.7,1);
 \end{tikzpicture}}
\ee
with superpotential
\be
\begin{split}
	W=\Tr\bigl[&X_{12}X_{24}X_{45}X_{51}-X_{24}X_{46}X_{62}+X_{23}X_{35}X_{56}X_{62}\bigr.\\
	&\bigl.-X_{35}X_{51}X_{13}+X_{34}X_{46}X_{61}X_{13}-X_{12}X_{23}X_{34}X_{45}X_{56}			X_{61}\bigr],
\end{split}
\ee
where $X_{ij}$ denotes a bifundamental field connecting the $i$-th and $j$-th nodes.
The charges assigned from the toric diagram to the various fields in the theory are
\be
\begin{array}{c|cccccc}
 &U(1)_1 & U(1)_2 & U(1)_3& U(1)_4& U(1)_5 &U(1)_R\\
 \hline
X_{12} & -1 & -1 & -1 & -1 & -1 & 1/3 \\
X_{13} & 0 & 1 & 1 & 0 & 0 & 2/3 \\
X_{23} & 0 & 0 & 0 & 0 & 1 & 1/3 \\
X_{24} & 1 & 1 & 0 & 0 & 0 & 2/3 \\
X_{34} & 0 & 0 & 0 & 1 & 0 & 1/3 \\
X_{35} & 0 & -1 & -1 & -1 & -1 & 2/3 \\
X_{45} & 0 & 0 & 1 & 0 & 0 & 1/3 \\
X_{46} & -1 & -1 & -1 & -1 & 0 & 2/3 \\
X_{56} & 0 & 1 & 0 & 0 & 0 & 1/3 \\
X_{51} & 0 & 0 & 0 & 1 & 1 & 2/3 \\
X_{61} & 1 & 0 & 0 & 0 & 0 & 1/3 \\
X_{62} & 0 & 0 & 1 & 1 & 0 & 2/3 \\
\end{array}
\ee
The entropy function for this theory is
\be
\label{dP3final}
\begin{split}
	S_E(\D,\o)=-\frac{i \pi N^2}{\o_1\o_2}\bigl(& \D_1\D_2\D_3+2\D_1\D_2\D_4 +2\D_1\D_3\D_4 
	+\D_2\D_3\D_4+2\D_1\D_2\D_5  \bigr. \\
	&\bigl.+3\D_1\D_3\D_5+2\D_2\D_3\D_5+2\D_1\D_4\D_5+2\D_2\D_4\D_5+\D_3\D_4\D_5   \bigr. \\
	&\bigl.+\D_1\D_2\D_6+2\D_1\D_3\D_6+2\D_2\D_3\D_6+2\D_1\D_4\D_6+3\D_2\D_4\D_6   \bigr. \\
	&\bigl.+2\D_3\D_4\D_6+\D_1\D_5\D_6+2\D_2\D_5\D_6+2\D_3\D_5\D_6+\D_4\D_5\D_6   \bigr).
\end{split}
\ee

In fact, evaluating $V$ at the leading order in $|\o_1|,|\o_2|\rightarrow 0$ we get
\be\label{VleaddP3}
\begin{split}
	V_{\text{leading}}=-\frac{i \pi N^2}{\o_1\o_2}\bigl(& 2 \D_3\D_4(1\!+\!\o_1\!+\!\o_2\!-\!\D_3\!-\!\D_4) \!-\!2 \D_3^2 \D_5\!+\!2 \D_3\D_5
	(1\!+\!\o_1\!+\!\o_2\!-\!2\D_4)\bigr. \\
	&\bigl. +\D_4\D_5(1\!+\! \o_1\!+\!\o_2\!-\!\D_4) \!+\!\D_5^2(2\D_3\!+\!\D_4)\!\!-\!\!\D_1^2(\D_2\!+\!2\D_3\!+\!2\D_4\!+\!\D_5)\bigr. \\
	&\bigl. -\D_2^2(2\D_3+3\D_4+2\D_5)-2\D_2\D_3^2+3\D_2\D_4(1+\o_1+\o_2-\D_4) \bigr. \\
	&\bigl. + 2\D_2\D_3(1+\o_1+\o_2-3 \D_4-2\D_5)+2 \D_2\D_5(1+\o_1+\o_2-2\D_4)\bigr. \\
	&\bigl.-2\D_2\D_5^2	- \D_1\D_2^2-2\D_1(\D_3+\D_4)(\D_3+\D_4-1-\o_1-\o_2)\bigr. \\
	&\bigl.+\D_1\D_2(1+\o_1+\o_2-4\D_3-4\D_4-2\D_5) -\D_1\D_5^2\bigr.\\
	&\bigl. +\D_1\D_5(1+\o_1+\o_2-2\D_3-2\D_4)\bigr),
\end{split}
\ee
where we also shifted the fugacities by $-\frac{\omega_1+\omega_2}{6}$, consistently with the general prescription.
Again, by imposing the constraint 
\be
	\D_6=-\D_1-\D_2-\D_3-\D_4-\D_5+\o_1+\o_2+1,
\ee
on \eqref{VleaddP3} we obtain \eqref{dP3final}.

\subsection{$\text{(P)dP}_4$}
The fourth del Pezzo surface is defined as the blow-up of $\mathbb{P}^2$ at four generic\footnote{Meaning that none of the possible triples of points lies on a line.} points. The (complex) cone over it possesses a Calabi-Yau structure and the theory living on $N$ D3-branes probing the conical singularity is known; however the superpotential of the dual gauge theory is such that no non-anomalous flavour symmetries except $U(1)_R$ are admitted, so that the model is non-toric. Blowing-up $\mathbb{P}^2$ at non-generic points, however, it is possible to build models where more symmetries are preserved.

One choice can be the toric model whose diagram is generated by the following vectors:
\be
\begin{split}
&V_1\,=\,(0,0,1)\,, V_2\,=\,(1,0,1)\,, V_3\,=\,(2,0,1)\,, V_4\,=\,(2,1,1)\,, \\
&V_5\,=\,(1,2,1)\,, V_6\,=\,(0,2,1)\,, V_7\,=\,(0,1,1)\,,
\end{split}
\ee
and that we will denote as pseudo $\text{dP}_4$ or $\text{(P)dP}_4$. The dual gauge theory can be described by the following quiver:
\be 
\scalebox{1.2}
{
\begin{tikzpicture}[font=\scriptsize, baseline] \begin{scope}[auto,%
  every node/.style={draw}, node distance=1cm]; 

 \node[draw=none] at (0,0.4) {$\bullet$};
 \node[draw=none] at (0,0.6) {1};

 \node[draw=none] at (-0.9,-0.2) {$\bullet$};
  \node[draw=none] at (-1.2,-0.2) {$2$};
   \node[draw=none] at (0.9,-0.2) {$\bullet$};
  \node[draw=none] at (1.15,-0.2) {$7$};
  
   \node[draw=none] at (-1.2,-1.2) {$\bullet$};
  \node[draw=none] at (-1.5,-1.2) {$6$};
   \node[draw=none] at (1.2,-1.2) {$\bullet$};
  \node[draw=none] at (1.5,-1.2) {$3$};

\node[draw=none] at (-0.7,-1.8) {$\bullet$};
  \node[draw=none] at (-1,-1.8) {$5$};
   \node[draw=none] at (0.7,-1.8) {$\bullet$};
  \node[draw=none] at (1.,-1.8) {$4$};
 \end{scope}
\draw[sdirected](0,0.4)-- (0.8,-0.2);
\draw[sdirected](0.9,-0.2)-- (1.2,-1.2);
\draw[sdirected](1.2,-1.2)--(-0.7,-1.8);
\draw[sdirected](0.7,-1.8)--(-1.2,-1.2);
\draw[sdirected](-1.2,-1.2)--(-0.9,-0.2);
\draw[sdirected] (-0.9,-0.2)--(0,0.4);
\draw[sdirected](0.7,-1.8)--(-0.7,-1.8);
\draw[sdirected](1.2,-1.2)--(-1.2,-1.2);
\draw[sdirected](-0.9,-0.2)--(0.9,-0.2);
\draw[sdirected](-1.2,-1.2)--(0,0.4);
\draw[sdirected](0,0.4)--(0.7,-1.8);
\draw[sdirected](0,0.4)--(1.2,-1.2);
\draw[sdirected](-0.7,-1.8)--(0,0.4);
\draw[sdirected](-0.7,-1.8)--(-0.9,-0.2);
\draw[sdirected](0.9,-0.2)--(0.7,-1.8);
 %\draw[sdirected]  (1.5,0) --(0.05,0);
      % \draw[directed] (0,0) -- (3,3);
 \end{tikzpicture}}
\ee
with superpotential:
\be
\begin{split}
W\,=\,\text{Tr}&\left[ X_{61}X_{17} X_{74}X_{46}+X_{21}X_{13}X_{35}X_{52}+X_{27}X_{73}X_{36}X_{62}+X_{14}X_{45}X_{51}\right.+\\
&\,\,\,\,-X_{51}X_{17}X_{73}X_{35}-X_{21}X_{14}X_{46}X_{62}-X_{27}X_{74}X_{45}X_{52}-X_{13}X_{36}X_{61}\left.\right]\,,
\end{split}
\ee
where each $X_{ij}$ must be understood as a field transforming in the bifundamental representation with respect to the $i$-th and $j$-th nodes. The charges of the fields can be parameterized in terms of the PMs using the assignation
\begin{equation}
\begin{split}
\begin{array}{c|c|c|c|c|c|c|c}
X_{17} & X_{21} & X_{27} & X_{73} & X_{14} &X_{74} & X_{13} & X_{62} \\
\hline
{\bf a_1+a_6+a_7}&{\bf a_7}&{\bf a_3+a_4}&{\bf a_2}&{\bf a_1+a_2+a_3}&{\bf a_5}&{\bf a_4+a_5+a_6}&{\bf a_5+a_6}
\end{array}\\
\begin{array}{c|c|c|c|c|c|c}
 X_{51} & X_{61} & X_{52} & X_{36} &X_{45} & X_{46} & X_{35}\\
\hline
{\bf a_4+a_5}&{\bf a_2+a_3}&{\bf a_1+a_2}&{\bf a_1+a_7}&{\bf a_6+a_7} &{\bf a_4} &{\bf a_3}
\end{array}
\quad\quad\quad\quad
\end{split}
\end{equation}
Thus, the  set of charges suitable for the underlying geometry is:
\be
\begin{array}{c|ccccccc}
&U(1)_1 & U(1)_2 & U(1)_3& U(1)_4& U(1)_5& U(1)_6 &U(1)_R\\
 \hline
X_{17} & 0 & -1 & -1 & -1 & -1&0 & 6/7 \\
X_{21} & -1& -1 & -1 & -1 & -1&-1 & 2/7 \\
X_{27} & 0& 0 & 1 &1 & 0&0 & 4/7 \\
X_{73} & 0& 1 & 0 &0 & 0&0 & 2/7 \\
X_{14} & 1& 1 & 1 &0 & 0&0 & 6/7 \\
X_{74} & 0& 0 & 0 &0 & 1&0 & 2/7 \\
X_{13} & 0& 0 & 0 &1 & 1&1 & 6/7 \\
X_{62} & 0& 0 & 0 &0 & 1&1 & 4/7 \\
X_{51} & 0& 0 & 0 &1 & 1&0 & 4/7 \\
X_{61} & 0& 1& 1 &0 & 0&0 & 4/7 \\
X_{52} & 1& 1& 0 &0 & 0&0 & 4/7 \\
X_{36} & 0& -1& -1 &-1 & -1&-1 & 4/7 \\
X_{45} & -1& -1& -1 &-1 & -1&0 & 4/7 \\
X_{46} & 0& 0& 0 &1 & 0&0 & 2/7 \\
X_{46} & 0& 0& 1 &0& 0&0 & 2/7 
\end{array}
\ee
We assign a fugacity $\D_i$ to each global $U(1)_i$ and, assuming $V_2$ has a local minimum for vanishing holonomies, we computed the entropy function following the same guide-line as before. By considering the $|\omega|_1,|\omega|_2\rightarrow 0$ and shifting each fugacity by $-\frac{\omega_1+\omega_2}{7}$ we obtain the following expression:
\be
\begin{split}
S_E\,=\, -\frac{i \pi\,N^2}{\omega_1\omega_2}&\left( \Delta_1 \Delta_2 \Delta_3+2\Delta_1 \Delta_2 \Delta_4+2\Delta_1\Delta_3\Delta_4+\Delta_2\Delta_3\Delta_4+2\Delta_1\Delta_2\Delta_5 \right.+\\
&\,\,\left.+2 \Delta_2\Delta_3\Delta_5+2\Delta_1\Delta_4\Delta_5+2\Delta_2\Delta_4\Delta_5+\Delta_3\Delta_4\Delta_5+\Delta_1\Delta_2\Delta_6\right.+\\
&\,\,\left.+2\Delta_1\Delta_3\Delta_6+2\Delta_2\Delta_3\Delta_6 +2\Delta_1\Delta_4\Delta_6+3\Delta_2\Delta_4\Delta_6+2\Delta_3\Delta_4\Delta_6\right.+\\
&\,\,\left.+\Delta_1\Delta_3\Delta_7+2\Delta_2\Delta_3\Delta_7+2\Delta_1\Delta_4\Delta_7+4\Delta_2\Delta_4\Delta_7+3\Delta_3\Delta_4\Delta_7\right.+\\
&\,\,\left.+2\Delta_1\Delta_5\Delta_7+4\Delta_2\Delta_5\Delta_7+4\Delta_3\Delta_5\Delta_7+2\Delta_4\Delta_5\Delta_7+\Delta_1\Delta_6\Delta_7\right.+\\
&\,\,\left.+3\Delta_1\Delta_3\Delta_5+\Delta_1\Delta_5\Delta_6+2\Delta_2\Delta_5\Delta_6+2\Delta_3\Delta_5\Delta_6+\Delta_4\Delta_5\Delta_6\right.+\\
&\,\,\,\left.+2\Delta_2\Delta_6\Delta_7+2\Delta_3\Delta_6\Delta_7+\Delta_4\Delta_6\Delta_7\right)\,.
\end{split}
\ee 
This is again in agreement with \eqref{expected}.

\section{Infinite families}
\label{Sec:Families}

In this section we compute the Cardy-like like limit of the superconformal index at large $N$ with complex fugacities,
for infinite families  of quiver gauge theories.
We assume that  the fugacities are in the regime 
$0 \leq \Delta_1,\dots,\Delta_{d-1} \leq \frac{1}{2}$  and  
$0 \leq \sum_{i=1}^{d-1} \Delta_i \leq 1$.  
In this regime we assume the validity of the conjecture on the existence of a universal saddle point
associated to the vanishing of the  holonomies. 
For each family we extract the entropy function and we verify the validity of the relation
(\ref{expected}).

\subsection{$Y^{pq}$}

We start our analysis with the $Y^{pq}$ quiver gauge theories,
introduced in \cite{Benvenuti:2004dy}.
They correspond to quiver gauge theories with $2p$ gauge groups and 
a chiral field content of  bifundamental fields.
When $p$ and $q$ are generic the models enjoy an $SU(2) \times U(1)$ flavor symmetry and 
in addition one non-anomalous $U(1)_B$.

The toric diagram is parameterized by the four vectors
\begin{equation}
V_1 = (0,0,1),\quad V_2 =(1,0,1),\quad V_3 =(0,p,1) \quad V_4 =(-1,p-q,1)
\end{equation}
As discussed in section \ref{Sec:Cardy} we can parameterize the global symmetries
using the toric data.
In this case there are four perfect matchings associated to the four external points
of the toric diagram.  The charges of the fields can be parameterized in terms of the PMs using the assignation
\begin{equation}
\begin{array}{c|c|c|c|c|c}
Y      &       U_1 &Z & U_2 & V_1 & V_2\\
\hline
{\bf a_1}&{\bf a_2}&{\bf a_3}&{\bf a_4}&{\bf a_2 +  a_3}&{\bf a_3+ a_4}
\end{array}
\end{equation}
we can use this parameterization to construct the basis of $U(1)_i$ symmetries that 
we will use in the calculation of the index.
Following the discussion in section \ref{Sec:Cardy} we have
\begin{equation}
\label{tableY}
\begin{array}{c|c|cccc}
   &\text{multiplicity} &U(1)_1 & U(1)_2 & U(1)_3 &U(1)_R\\
\hline
Y         &p+q&  1&0&0 & \frac{1}{2} \\ 
U_1     &p& 0&1&0  &\frac{1}{2} \\
Z         &p-q& 0 &0 &1 & \frac{1}{2} \\
U_2     &p& -1 &-1 &-1& \frac{1}{2} \\ 
V_1     &q& -1 &0 &0  & 1\\
V_2     &q& -1 &-1 &0&1\\ 
\end{array}
\end{equation}
We can assign a  fugacity  $\Delta_i$ to th $i$-th $U(1)$ in this table.
Furthermore we assign an equal $R$-symmetry to each perfect matching, such that 
the $R$-charges of the fields are given in the table.
Then we shift each fugacity by $-\frac{\omega_1+\omega_2}{4}$, where $4$ refers to the number 
of points in the toric diagram.
%We also assign a trial R-symmetry to each field as in the table above.
After this shift we compute the Cardy-like limit of the index at the universal saddle point, {\it i.e.} by setting all the gauge holonomies 
to zero. In this way, at large $N$, the leading contribution to the index, corresponding to the entropy function
\begin{eqnarray}
\label{SEYpq}
S_E = 
-\frac{i \pi N^2} {\omega_1\omega_2}
&\Big(&
p \Delta_1 \Delta_2 \Delta_3  + ((p+q) \Delta_1 \Delta_2 + p \Delta_1 \Delta_3 +
 \nonumber \\
&+&
(p-q) \Delta_2 \Delta_3) (1+\omega_1+\omega_2-\Delta_1-\Delta_2-\Delta_3)
\Big)  
\end{eqnarray}
Defining $\Delta_4 \equiv 1+\omega_1+\omega_2-\Delta_1-\Delta_2-\Delta_3$
the entropy function in (\ref{SEYpq}) becomes 
\begin{eqnarray}
S_E &=& 
-\frac{i \pi N^2}{\omega_1\omega_2}
\Big(
p \Delta_1 \Delta_2 \Delta_3  + (p+q) \Delta_1 \Delta_2 \Delta_4 + p \Delta_1 \Delta_3 \Delta_4 +(p-q) \Delta_2 \Delta_3 \Delta_4 \Big)
\nonumber\\
\end{eqnarray}
It is straightforward to check that 
the final form of the entropy function is then given by (\ref{expected}),
where the coefficients $C_{IJK}$, are computed from  (\ref{BPT}).

\subsection{$X^{pq}$}      

These models have been introduced in \cite{Hanany:2005hq}. For generic values of $p$ and $q$ there are $2p + 1$ gauge groups, there is a $U(1)^2$
flavor symmetry and two non anomalous baryonic $U(1)$ symmetries.
The toric diagram is parameterized by the five vectors
\begin{equation}
V_1 = (1-q,1,1),  \, V_2 =(-1,0,1),  \,  V_3 =(q-p,-1,1), \, V_4 =(0,-1,1),
 \, V_5 =(p,1,1)
\end{equation}
As discussed in section \ref{Sec:Cardy} we can parameterize the global charges
as
\begin{equation}
\label{tableX}
\begin{array}{c|ccccc}
 \text{multiplicity} &U(1)_1 & U(1)_2 & U(1)_3 & U(1)_4 &U(1)_R\\
 \hline
 p+q-1 & 1 & 0 & 0 & 0 & \frac{2}{5} \\
 1 & 0 & 1 & 0 & 0 & \frac{2}{5} \\
 1 & 0 & 0 & 1 & 0 & \frac{2}{5} \\
 p-q & 0 & 0 & 0 & 1 & \frac{2}{5} \\
 p & -1 & -1 & -1 & -1 & \frac{2}{5} \\
 p-1 & 0 & 1 & 1 & 0 & \frac{4}{5} \\
 1 & 0 & 0 & 1 & 1 & \frac{4}{5} \\
 q-1 & 0 & 1 & 1 & 1 & \frac{6}{5} \\
 1 & 1 & 1 & 0 & 0 & \frac{4}{5} \\
 q & -1 & -1 & -1 & 0 & \frac{4}{5} \\
\end{array}
\end{equation}
Then we shift each charge by $-\frac{\omega_1+\omega_2}{5}$, where $5$ refers to the number 
of points in the toric diagram.
We also assign a trial R-symmetry to each field as in the table above.
Then we compute the Cardy-like limit of the index by setting all the holonomies
to zero and supposing that there exists a regime of charges such that
a minimum exists. In this way, at large $N$, the entropy function is 
\begin{eqnarray}
S_E = -
\frac{i \pi N^2}{\omega_1\omega_2}
&\Big(&
\Delta_1 \Delta_5 \Delta_3 (p+q)+\Delta_4 \Delta_5 \Delta_3 (p-q)+
\Delta_1 \Delta_2 \Delta_5 (p+q-1)
\nonumber \\
&+& 
\Delta_2 \Delta_4 \Delta_5 (p-q+1)
+\Delta_1 \Delta_4 \Delta_5 p+
\Delta_1 \Delta_4 \Delta_3 p+\Delta_1 \Delta_2 \Delta_4 p
\nonumber \\
&+& 
\Delta_1 \Delta_2 \Delta_3
+\Delta_2 \Delta_4 \Delta_3+2 \Delta_2 \Delta_5 \Delta_3
\Big)
\end{eqnarray}
where we defined $\Delta_5 = 1+\omega_1+\omega_2-\Delta_1-\Delta_2-\Delta_3-\Delta_4$.

This formula can be interpreted in terms of the geometric data by assigning the charges 
$\Delta_I$ the four corners of the toric diagram.
The final form of the entropy function is then given by (\ref{expected}),
where the coefficients $C_{IJK}$ are computed from  (\ref{BPT}).

\subsection{$L^{pqr}$}   
   
   These models have been introduced in \cite{Benvenuti:2005ja,Butti:2005sw,Franco:2005sm}.
The toric diagram is parameterized by the four vectors
\begin{equation}
V_1 = (0,0,1),\quad V_2 =(1,0,1),\quad V_3 =(P,s,1) \quad V_4 =(-k,q,1)
\end{equation}
where $P q = r-ks$.
If $p \neq r$ 
we can parameterize the global charges
as
\footnote{
In the case $p=r$ the toric diagram gains a large amount of external points 
 lying on the perimeter. It induces a large set of non-anomalous baryonic
 symmetries in the quiver.}
\begin{equation}
\label{tableL}
\begin{array}{c|c|cccc}
   &\text{multiplicity} &U(1)_1 & U(1)_2 & U(1)_3 &U(1)_R\\
\hline
Y         &q&  1&0&0 & \frac{1}{2} \\ 
W_2     &s& 0&1&0  &\frac{1}{2} \\
Z         &p& 0 &0 &1 & \frac{1}{2} \\
X_2     &r& -1 &-1 &-1&\frac{1}{2}\\ 
W_1   &q-s& 0 &1 &1& 1 \\ 
X_1    &q-r& -1 &-1 &0  & 1\\
\end{array}
\end{equation}
where $p+q=r+s$.

Then we shift each charge by $-\frac{\omega_1+\omega_2}{4}$, where $4$ refers to the number 
of points in the toric diagram.
We also assign a trial $R$-symmetry to each field as in the table above.
Then we compute the Cardy-like limit of the index by setting all the holonomies
to zero and supposing that there exists a regime of charges such that
a minimum exists. In this way, at large $N$, we can show that the entropy function is 
equivalent to 
\begin{eqnarray}
S_E &=& 
-\frac{i \pi N^2}{\omega_1\omega_2} \Big(
 s \Delta_1 \Delta_2 \Delta_3 +q \Delta_1 \Delta_2 \Delta_4 + r \Delta_1 \Delta_3 \Delta_4 +p \Delta_2 \Delta_4 \Delta_3
 \Big)
\end{eqnarray}
where $\Delta_4 = 1+\omega_1+\omega_2-\Delta_1-\Delta_2-\Delta_3$.

This formula can be interpreted in terms of the geometric data by assigning the charges 
$\Delta_I$ the four corners of the toric diagram.
The final form of the entropy function is then given by (\ref{expected}),
where the coefficients $C_{IJK}$ are computed from  (\ref{BPT}).

\section{Legendre transform and the entropy}
\label{Sec:Legendre}

In this section we obtain the entropy associated to some of the families discussed above.
We focus on the $Y^{pp}$ and on the $L^{0b0}$ cases.
These two cases are similar to the $\mathcal{N}=4$ case because they can 
be constructed by an orbifold projection of $\mathbb{C}^3$.
At the level of the toric diagram this reflects in the fact that there are three corners.
The other external points are on the perimeter, signaling the presence of 
 non smooth horizons induced by the orbifold.
These models are anyway richer, because they have a higher amount of global, baryonic, symmetries.
In this section we compute the Legendre transform of the entropy function for these models, by 
turning on all of the possible global symmetries.
The ones discussed in this section are the only cases where we have found an expression for the 
entropy by computing the Legendre transform.
For other geometries with more then three corners in the toric diagram 
and all the global symmetries  turned on, we have not found a systematic way
to compute the Legendre transform of the entropy function.

Let us also comment on the $C_{IJK}$ coefficients for the theories that 
we discuss below with respect to the 
multi-charge AdS$_5$ black holes  obtained from gauged supergravity in \cite{Kunduri:2006ek}. 
On the supergravity side the condition 
\be
\label{condsym}
	C_{IJK}C_{J'(LM}C_{PQ)K'}\delta^{JJ'}\delta^{KK'}=\frac{4}{3}\delta_{I(L}C_{MPQ)}
\ee
was imposed, while here we have explicitly checked that the $C_{IJK}$ coefficients
discussed in this section do not satisfy (\ref{condsym}).

\subsection{The $Y^{pp}$ family}

In this section we study the Legendre transform of the entropy function obtained in 
the case of $Y^{pp}$ models.
The entropy function is given by 
\be
\label{SEYPP}
S_E = -\frac{i \pi p N^2}{\omega_1 \omega_2}(\Delta_1 \Delta_2 \Delta_3+\Delta_1 \Delta_3 \Delta_4 + 2 \Delta_1 \Delta_2 \Delta_4)
\ee
with $\sum_I \Delta_I = \omega_1 + \omega_2-1$.
The Legendre transform is computed in terms of the conjugate charges $Q_I$  and angular momenta $J_a$ and it corresponds to 
\be
S(Q,J) =S_E (X,\omega)+2\pi i (\sum_{I=1}^{4} \Delta_I Q_I + \sum_{a=1}^{2} \omega_a J_a)
+2 \pi i \Lambda  (\sum_{I=1}^{4} \Delta_I  - \sum_{a=1}^{2} \omega_a -1)
\ee
Observing that 
\be
S_E (X,\omega) = \left(\sum_{I=1}^{4} \Delta_I  \frac{\partial S_E}{\partial \Delta_I} + \sum_{a=1}^{2} \omega_a  \frac{\partial S_E}{\partial \omega_a} \right)
\ee
we have that $S(Q,J) = 2 \pi i \Lambda$.
The Lagrange multiplier can be obtained from the equation
\begin{eqnarray}
&&
(\Lambda +Q_1) 
(
2 (2 (\Lambda +Q_3)+(\Lambda +Q_4)) (\Lambda +Q_2)
\!-\!(\Lambda +Q_2)^2
\!-\!(2 (\Lambda +Q_3)\!-\!(\Lambda +Q_4))^2
)
\nonumber \\
&&
+4 N^2 p \left(\Lambda -J_1\right) \left(\Lambda -J_2\right) = 0
\end{eqnarray}
Reorganizing the polynomial on the LHS of this equation in the form 
$\Lambda^3+\Lambda^2 p_2 + \Lambda p_1 + p_0$ we have two imaginary solutions 
if $p_0 = p_1 p_2$.
The coefficients $p_i$ in this case are 
\begin{eqnarray}
p_2&=&N^2 p+Q_1+Q_2+Q_4
\nonumber \\
p_1&=&(Q_1 +Q_3)(Q_2+Q_4)+\frac{Q_4 Q_2}{2}-\frac{Q_2^2}{4}-Q_3^2-\frac{Q_4^2}{4}
-N^2 p (J_1 +J_2)
 \\
p_0&=&N^2 p J_1 J_2 -\frac{1}{4} Q_1 Q_2^2+Q_1 Q_3 Q_2+\frac{1}{2} Q_1 Q_4 Q_2-Q_1 Q_3^2-\frac{1}{4} Q_1 Q_4^2+Q_1 Q_3 Q_4
\nonumber 
\end{eqnarray}
and the entropy corresponds to  
\begin{eqnarray}
S(Q,J)  =2 \pi \sqrt{
(Q_1 +Q_3)(Q_2+Q_4)+\frac{Q_4 Q_2}{2}-\frac{Q_2^2}{4}-Q_3^2-\frac{Q_4^2}{4}
-N^2 p (J_1 +J_2)
}
\nonumber \\
\end{eqnarray}

Furthermore this model has been recently analyzed by \cite{Kim:2019yrz}.
The authors studied the entropy by turning off the fugacity for the 
$SU(2)_L$ symmetry. It corresponds here to turn off the 
variable $\Delta_2$ in (\ref{SEYPP}).
In this case the entropy function becomes 
\be
\label{SEYPP2}
S_E (X,\omega) =- \frac{i \pi p N^2}{\omega_1 \omega_2}(\Delta_1 \Delta_3 \Delta_4)
\ee
with the constraint $\Delta_1+\Delta_3 +\Delta_4 = \omega_1+\omega_2-1$.
One can repeat the analysis discussed above. 
The relevant  point of the discussion is that 
in this case the cubic equation for the Lagrange multiplier is
\be
(\Lambda+Q_1) (\Lambda+Q_3) (\Lambda+Q_4) 
+
\frac{N^2 p }{2} (\Lambda-J_1)(\Lambda-J_2) =0
\ee
and the entropy becomes 
\begin{eqnarray}
S(Q,J)  =2 \pi \sqrt{
Q_1 Q_3+Q_1 Q_4 + Q_3 Q_4 - \frac{N^2 p}{2}(J_1 +J_2 )
}
\end{eqnarray}
This result can be mapped to the one of \cite{Kim:2019yrz}
by mapping the charges $Q_I$ to the ones discussed there.

\subsection{The $L^{aba}$ family}

Here we compute the Legendre transform of the entropy function 
of a family of necklace quivers, that correspond to the $L^{0b0}$ family
studied above.
This class of models has already been discussed
by \cite{Honda:2019cio}, where it has been shown that the orbifold modifies just an overall contribution to the entropy function.
This was proven by just studying the effect of the flavor symmetries, but in this case
there are in addition $G-1$ non-anomalous baryonic symmetries, being $G$ the number
of gauge groups in the necklace.
Here we study the entropy function for a generic parameterization of the
charges, taking care of all the non anomalous baryonic symmetries as well.

As discussed in sub-section \ref{SPP} the entropy function can be expressed in terms 
of the areas of the toric diagram, contracted with the fugacities $\Delta_I$,
with $I=1,\dots,d$ runs over all the external points of the toric diagram.
In this case the formula can be expressed as
\begin{equation}
S_E =-i \pi N^2 \frac{\Delta_1\sum_{i,j=2}^{d} |i-j| \Delta_i \Delta_j}{2 \omega_1 \omega_2}
\end{equation}
The entropy is given by the Legendre transform
\begin{equation}
S(Q,J) = S_E(\Delta,\omega) + 2 \pi i (\sum_{I=1}^d \Delta_I Q_I +  \sum_{a=1}^{2} \omega_a J_a)
+ 
2 \pi i \Lambda (\sum_{I=1}^{d} \Delta_I -\sum_{a=1}^{2} \omega_a -1)
\end{equation}
The relation 
\begin{equation}
S_E = \sum_{I=1}^d \Delta_I \frac{\partial S_E}{\partial \Delta_I} +  \sum_{a=1}^{2} \omega_a
\frac{\partial S_E}{\partial J_a} 
\end{equation}
guarantees that
\begin{equation}
S_Q = 2 \pi i \Lambda
\end{equation}
By using the equations of motion we have induced a cubic  
relation satisfied by the Lagrange multiplier $\Lambda$. 
The cubic relation for $\Lambda$ is 
\be
\begin{split}
 	&(Q_{1}+\Lambda)	\left( \sum_{i=2}^{d-1}(Q_i+\Lambda)(Q_{i+1}+\Lambda)- \sum_{i=2}^{d-1}(Q_i+\Lambda)^2  +\frac{(Q_2+\Lambda)(Q_d+\Lambda)}{d-2} \right. \\
	&\left. +\frac{(d-1)(\left(Q_2+\Lambda)^2+(Q_d+\Lambda)^2\right)}{2(d-2)}  \right)+N^2(\Lambda-J_1)(\Lambda-J_2)=0
\end{split}
\ee
This equation is of the form $\Lambda^3+p_2\Lambda^2+p_1\Lambda +p_0=0$, with
\be
\begin{split}
	p_2=& (d-2)\frac{N^2}{2}+Q_1+Q_2+Q_d,\\
	p_1=&2 Q_1(Q_2+Q_d)+Q_2Q_d
	-\frac{d-1}{2} (Q_2^2 +Q_d^2)  \\
	&
	+(d-2)\bigg( \sum_{i=2}^{d-1}Q_i Q_{i+1}-\sum_{i=2}^d Q_i^2 \bigg)
	-(d-2)\frac{N^2}{2}(J_1+J_2)\\
	p_0=&(d-2)\frac{N^2}{2} J_1J_2 +\frac{1}{4}Q_1\left( (d-1)(Q_2^2+Q_d^2)+2Q_2Q_d\right.  \\
	&\left.-2(d-2)\sum_{i=2}^dQ_i^2 +2(d-2) \sum_{i=2}^{d-1}Q_iQ_{i+1}\right).
\end{split}
\ee
There is an imaginary solution for $p_2 p_1 = p_0$, and in this case the entropy is given by 
$\Lambda = - i \sqrt p_1$, or more explicitly
\be	
\begin{split}
	S(Q,J)=&2\pi \left(2 Q_1(Q_2+Q_d)+Q_2Q_d
	-\frac{d-1}{2} (Q_2^2 +Q_d^2) \right. \\
	&\left. 
	+(d-2)\bigg( \sum_{i=2}^{d-1}Q_i Q_{i+1}-\sum_{i=2}^d Q_i^2 \bigg)
	-(d-2)\frac{N^2}{2}(J_1+J_2)\right)^{\tfrac{1}{2}}
\end{split}
\ee
Observe that such expression correctly gives us back the entropy for $\mathcal N=4$ SYM case for $d=3$.

\section{Conclusions}
\label{Sec:Conclusions}

In this paper we studied the Cardy-like behavior of the SCI in presence of 
complex fugacities. This quantity has been recently observed to 
reproduce, in the case of $\mathcal{N}=4$ SYM, 
the entropy of an AdS$_5$ rotating  black hole, through a Legendre transform.
Here we focused on infinite families of 4d $\mathcal{N}=1$ quiver gauge theories,
describing stacks of D3 branes probing the tip of a toric CY$_3$ cones
over a 5d SE$_5$ base.
We showed that the general formula (\ref{expected})
for the entropy function of the models under investigation 
can be obtained from the Cardy-like limit of the 
SCI with complex fugacities.
Furthermore we computed the Legendre transform for 
some of the models analyzed here, giving a prediction for the 
entropy of the dual black hole.

In the analysis we left many open questions that deserve further investigation.
First we conjectured that it is always possible to 
find a regime of charges such that the holonomies are vanishing at the saddle point.
This conjecture is consistent with the ones given in \cite{Choi:2018hmj,Honda:2019cio,ArabiArdehali:2019tdm}.
Further arguments in favor of this idea has been recently given by \cite{Cabo-Bizet:2019osg}.
%It should be also interesting to discuss the other regimes of 
%couplings where by applying the conjecture we did not find a 
%saddle but a flat potential. 
In addition we have obtained the expected result in a regime of fugacities corresponding to 
the choices $0 \leq \Delta_1,\dots,\Delta_{d-1} \leq \frac{1}{2}$  and  
$0 \leq \sum_{i=1}^{d-1} \Delta_i \leq 1$.
In other regimes we have not found a minimum of the potential $V_2(a)$ but a plateau.
Similar plateaux have been discussed in \cite{Honda:2019cio}, but in that case they appeared for the choice
$\text{Re}\left( \frac{i}{\omega_1 \omega_2} \right)<0$ and they were associated to the Stokes lines discussed also in
\cite{Benini:2018ywd}. Here the plateaux  appear also in the regime 
$\text{Re} \left( \frac{i}{\omega_1 \omega_2} \right)>0$ and it should be interesting to have a deeper understanding of them 
and of their holographic dual interpretation.
Furthermore, we did not find a general way to obtain the Legendre transform
for entropy functions associated to toric diagram with more than three external corners
if all the global symmetries are turned on.
It should be interesting to see if this is just a technical obstruction or if is there a
deeper physical reason.
We conclude observing that a similar geometric relation between the Cardy-like limit of the SCI and the entropy function
can be expected for non-toric cases, as the one discussed in 
\cite{Butti:2006nk}. It should be interesting to investigate along this line.

\section*{Acknowledgments}

We are grateful to Alberto Zaffaroni, Alessandra Gnecchi, Francesco Benini and Noppadol Mekareeya  
for useful comments. This work has been supported
in part by Italian Ministero dell'Istruzione, Universit\`a e Ricerca (MIUR) and Istituto
Nazionale di Fisica Nucleare (INFN) through the ``Gauge Theories, Strings,
Supergravity" (GSS) research project. 
We gratefully acknowledge the ICTP where  some of
the research for this paper was performed during the 
2019 ``Spring School on Superstring Theory and Related Topics".
A.~A.~  thanks CERN for the
hospitality during the completion of this project.

% \newpage
 
\appendix
\section{Saddle point analysis for the Conifold at higher-rank }
\label{rank2Conifold}

In section \ref{Sec:conifold} we studied the behaviour of the minima of $V_2$ with respect to the fugacity range in the case of $SU(2)$ gauge groups. In this appendix we want to collect some evidence about the possibility of extending those considerations to higher ranks.
Let us briefly remind the main results: $V_2$ can be expressed as 
\begin{equation}
\label{V2conifold1}
\begin{split}
V_2\,=\,\!-\!\sum_{m,n=1}^{N} &\left(\kappa\left[a_m^{(1)}\!-\!a_n^{(2)}+\D_{F_1}+\D_{F_2}+\D_{B}\right]+\kappa\left[a_m^{(1)}-a_n^{(2)}\!-\!\D_{F_1}\!-\!\D_{F_2}+\D_{B}\right]+\right.\\ 
&\left.\,\,\,\,\kappa\left[a_m^{(2)}\!-\!a_n^{(1)}+\D_{F_1}\!-\!\D_{F_2}-\D_{B}\right]+\kappa\left[a_m^{(2)}-a_n^{(1)}\!-\!\D_{F_1}+\D_{F_2}\!-\!\D_{B}\right] \right)\,,
\end{split}
\end{equation}
where $a^{(k)}_{m}$ are holonomies for $k$-th gauge group and $\D_{F_1},\D_{F_2}, \D_B$ are fugacities for flavour and baryonic symmetries. At rank $N-1$ we have to enforce the constraint:
\be
a^{(k)}_{N}\,=\,-\sum_{m=1}^{N-1} a^{(k)}_{m}, \quad k=1,2\,,
\ee
so that in rank-1 case we are left with just two independent variables. We fixed a chamber where
\be
0 \leq \D_{F_1}\leq \D_{F_2} \leq \D_B\leq 1/2\,,\quad 0\leq \D_{F_1}+\D_{F_2}+\D_{B}\leq 1/2\,,
\ee
finding two possible behaviours for $V_2$:
\begin{itemize}
\item $\mathbf{\D_B> \D_{F_1}+\D_{F_2}}$: $V_2$ admits only plateaux of minima and thus the index is hard to evaluate.
\item $\mathbf{\D_B<\D_{F_1}+\D_{F_2}}$: $V_2$ admits a local maximum for vanishing holonomies that dominates in the Cardy-like limit.
\end{itemize}
Performing a similar analysis for higher rank is more complicated, because more variables are involved; however we can use the high symmetry of the model to simplify the computation: a natural expectation is that at high temperature, {\it i.e.} in the Cardy limit, all the global symmetries are preserved and no gauge symmetries are broken, so that no Higgs mechanisms are involved. In other words, we want to count  the degrees of freedom of the theory in the deconfining phase. A symmetry that we expect to be preserved at high temperature is a $\mathbb{Z}_2$ discrete symmetry of the quiver, exchanging the two nodes and the two couples of bifundamental fields; in order to keep this symmetry, we impose the following cyclic condition:
\be
\label{cyclic}
a^{(1)}_m\,=\,a^{(2)}_m\,\quad \forall \, m\,,
\ee
as already suggested in \cite{Honda:2019cio}. 

\begin{figure}[!h]
\centering
  \includegraphics[width=0.4\textwidth]{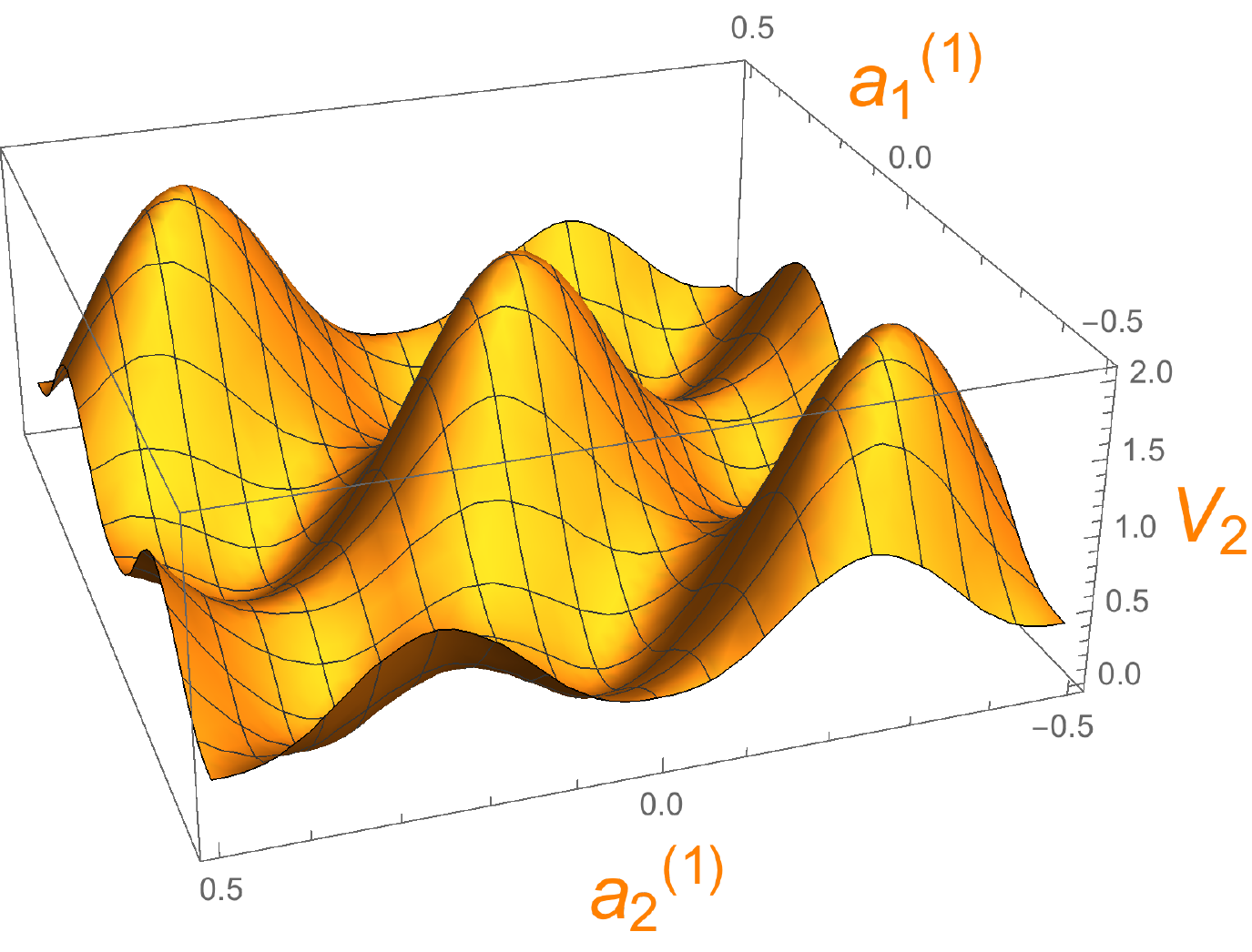}\quad  \includegraphics[width=0.4\textwidth]{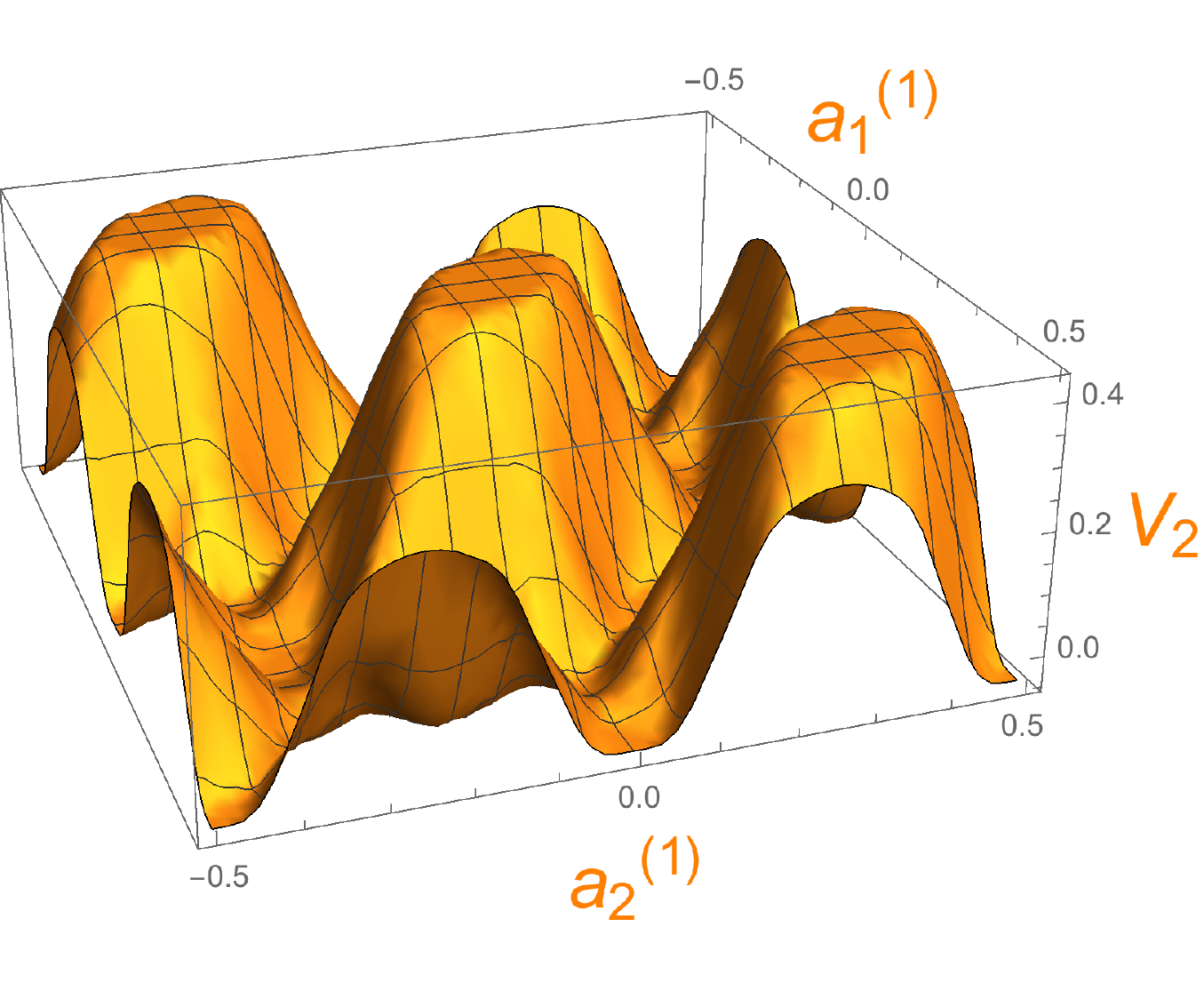}
    \caption{$V_2$ for $SU(3)$ conifold with cyclically identified holonomies: on the left we fixed $\D_{F_1}=0.12,\D_{F_2}=0.15, \D_B=0.22$ and we can observe that the function enjoys a local maximum at the origin; on the right we fixed $\D_{F_1}=0.05,\D_{F_2}=0.1, \D_B=0.3$ and we can observe that $V_2$ only possesses plateaux.}
    \label{ConifoldSU3}
    \end{figure}

For the rank-2 case this is enough in order to study numerically $V_2$, that is again a function of two variables only, $a^{(1)}_{1}$
and $a^{(1)}_{2}$. The plots of the function in figure \ref{ConifoldSU3} shows that $V_2$ still shares the same properties as before. 

\begin{figure}[H]
\centering
  \includegraphics[width=0.4\textwidth]{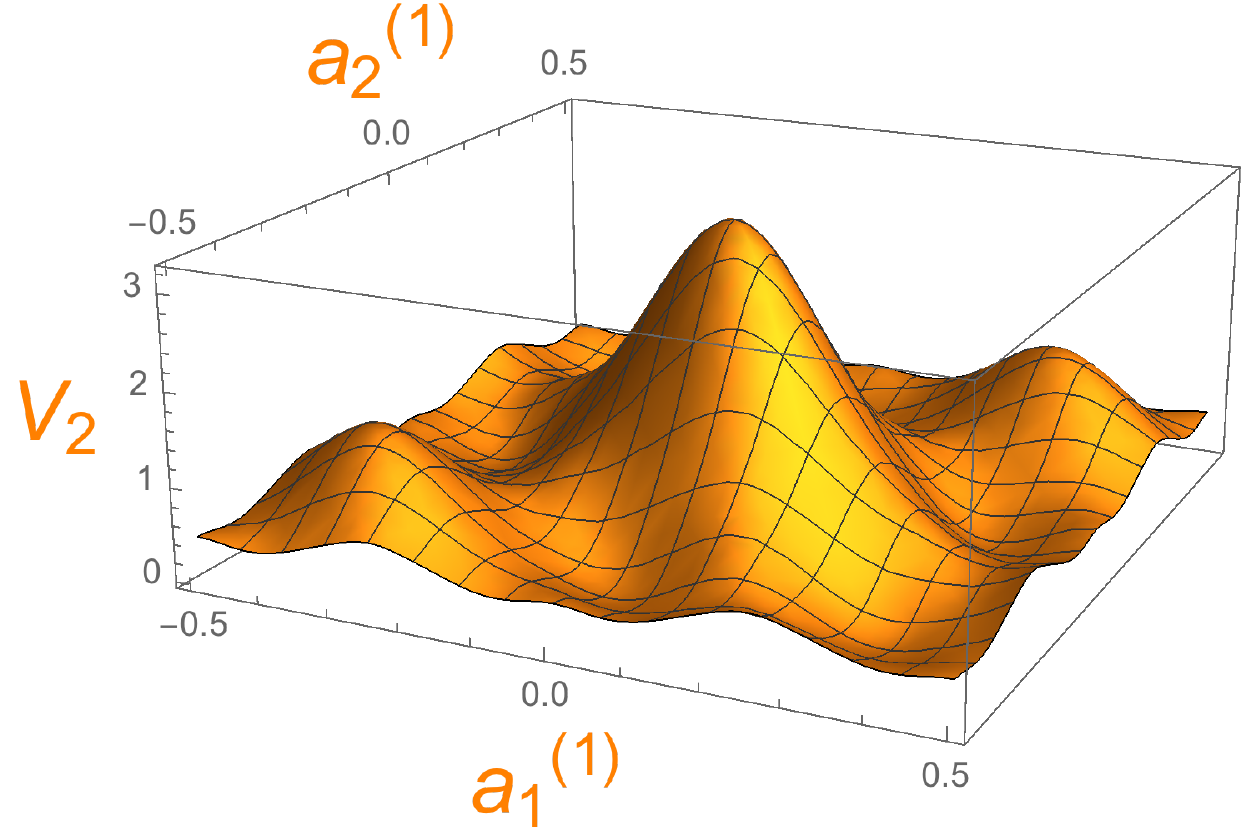}\quad  \includegraphics[width=0.4\textwidth]{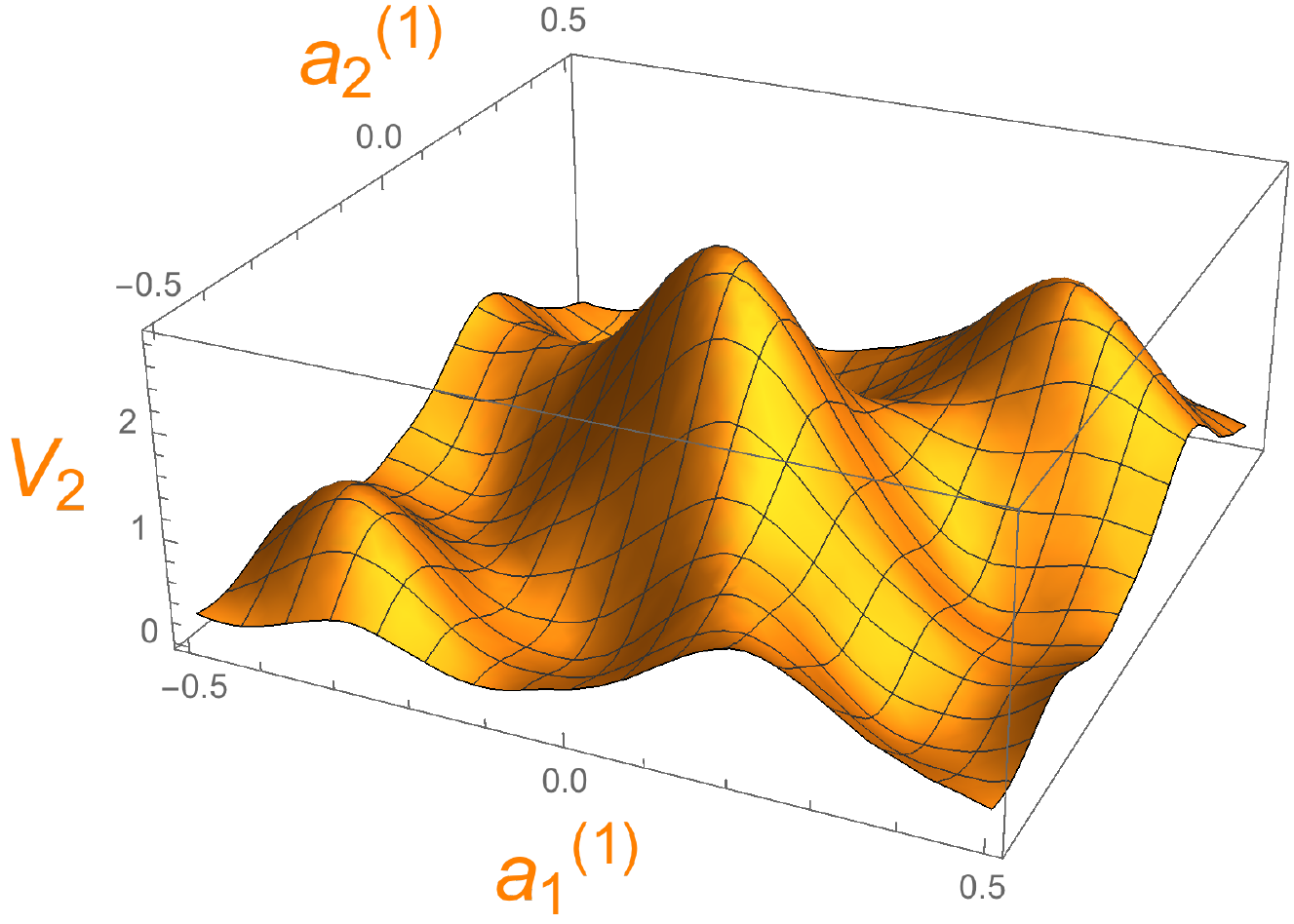}\quad  \includegraphics[width=0.4\textwidth]{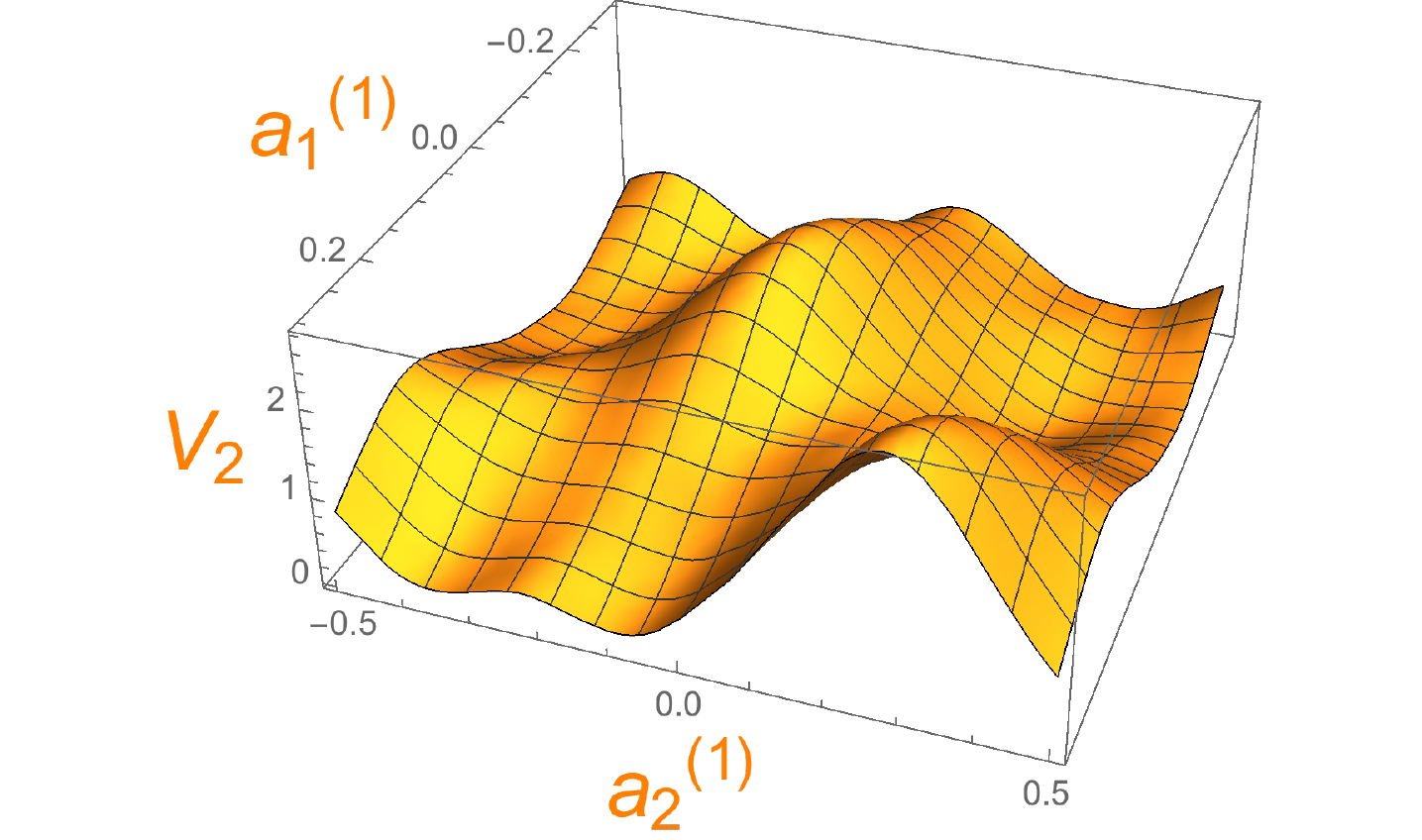}
    \caption{$V_2$ in the  $a^{(1)}_1-a^{(1)}_2$ plane for $SU(4)$ conifold with cyclically identified holonomies and $\D_{F_1}=0.1,\D_{F_2}=0.15, \D_{B}=0.2$; from the top left in clockwise sense we fixed $a^{(1)}_3=0$,  $a^{(1)}_3=0.08$  and  $a^{(1)}_3=0.16$. We notice a minimum whose lowest value is reached for $a^{(1)}_3=0$.}
    \label{ConifoldSU4Min}
    \end{figure}

\begin{figure}[!h]
\centering
  \includegraphics[width=0.4\textwidth]{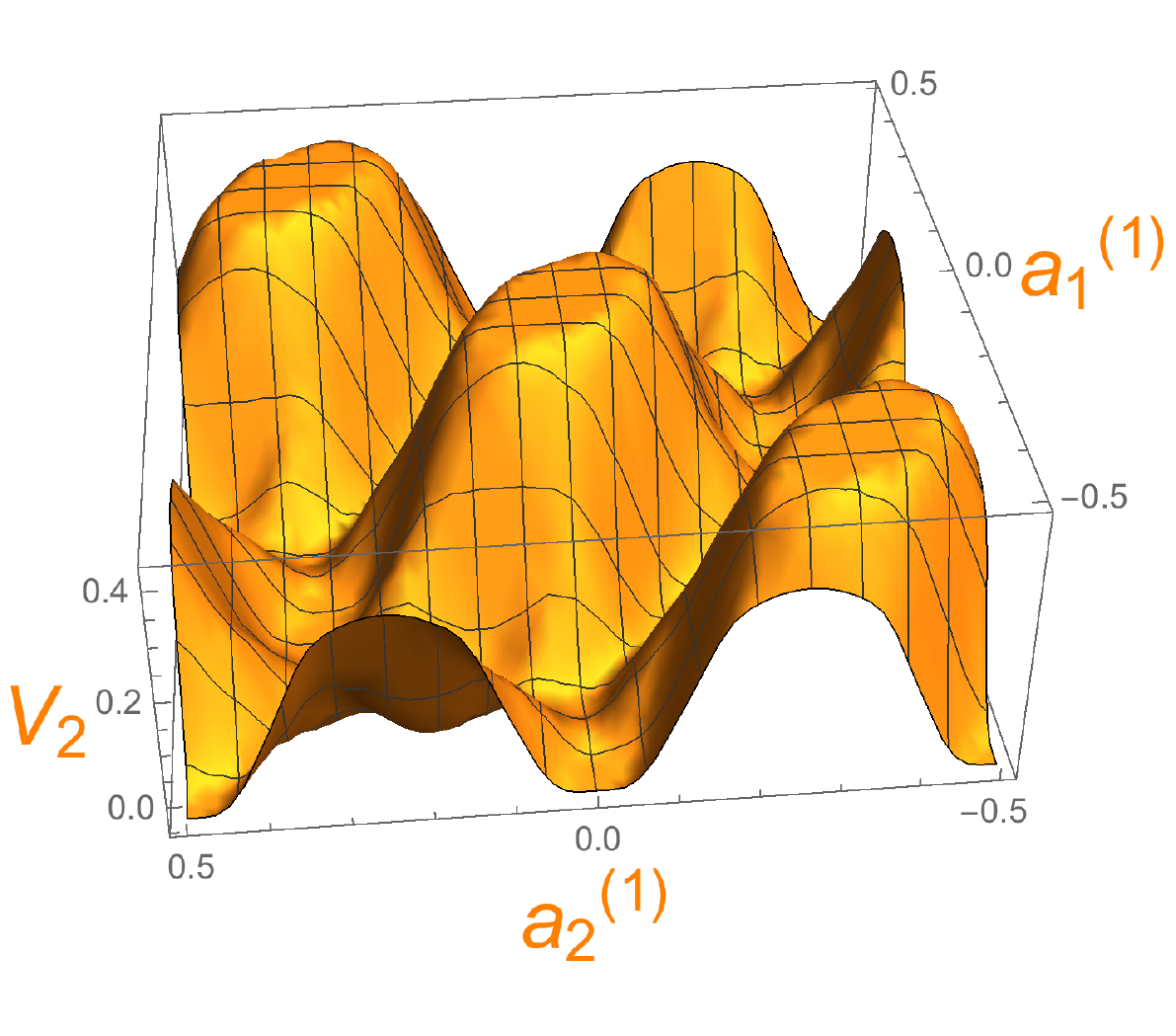}
    \caption{$V_2$ in the  $a^{(1)}_1-a^{(1)}_2$ plane for $SU(4)$ conifold with cyclically identified holonomies and $\D_{F_1}=0.05,\D_{F_2}=0.1, \D_{B}=0.3\,, a^{(1)}_3\,=\,0$; only plateaux are present.}
    \label{ConifoldSU4Plat}
    \end{figure}
    We can perform a similar analysis at rank 3. In this case we have three independent variables, $a^{(1)}_{1}$,
$a^{(1)}_{2}$ and $a^{(1)}_{3}$ and thus we cannot make a single plot; we need to use a slightly different technique: we can make a plot $V_2$ in the $a^{(1)}_1-a^{(1)}_2$ plane at fixed $a^{(1)}_3$ and then vary the value of this last holonomy. If a minimum is located at the origin, $V_2$ restricted to the $a^{(1)}_1-a^{(1)}_2$ plane should have a minimum, as deep as we get closer to $a^{(1)}_3\,=\,0$. This is the kind of behaviour that we can observe in figure \ref{ConifoldSU4Min}, where we fixed the fugacity in such a way the condition $\D_{F_1}+\D_{F_2}>\D_B$ to hold. When $\D_{F_1}+\D_{F_2}<\D_B$, instead, the presence of plateaux is already evident from a plot of $V_2$ at $a^{(1)}_3=0$, as shown in figure \ref{ConifoldSU4Plat}.
Let us stress finally stress that, even relaxing the assumption \eqref{cyclic}, we can use other numerical tools such as {\bf FindMaximum/FindMinimum} of Mathematica in order to understand the behaviour of $V_2$; this kind of study still returns the same results as before.

%\newpage
\bibliographystyle{ytphys}
\bibliography{ref}

\end{document}